\documentclass[aps, prd, twocolumn, superscriptaddress,nofootinbib]{revtex4}
\usepackage{graphicx}
\usepackage{dcolumn}
\usepackage{amssymb}

\begin{document}

\newcommand{\vect}[1]{\mathbf{#1}}
\newcommand{\half}{^{1}\!\!/_{\!2}}
\newcommand{\etal}{\emph{et al.\ }}
\newcommand{\conj}{^{*}}
\newcommand{\sY}{\sigma_{\mathrm{Y}}}
\newcommand{\sK}{\sigma_{\mathrm{K1}}}
\newcommand{\gammaU}{\gamma_{u}}
\newcommand{\gammaW}{\gamma_{w}}
\newcommand{\sdot}[1]{\dot{s}_{#1}}
 
\title{Cosmic string Y-junctions: a comparison between field theoretic and Nambu-Goto dynamics}

\newcommand{\addressSussex}{Department of Physics \&
Astronomy, University of Sussex, Brighton, BN1 9QH, United Kingdom}

\newcommand{\addressImperial}{Theoretical Physics, Blackett Laboratory, Imperial College, Prince Consort Road, London, SW7 2BZ, United Kingdom}

\newcommand{\addressNottingham}{School of Physics \&
Astronomy, University of Nottingham, University Park, Nottingham, NG7 2RD, United Kingdom}

\author{Neil Bevis} 
\email{n.bevis@imperial.ac.uk}
\affiliation{\addressImperial}

\author{Paul M. Saffin} 
\email{paul.saffin@nottingham.ac.uk}
\affiliation{\addressNottingham}

\date{3 July 2008}

\begin{abstract}
We explore the formation of cosmic string Y-junctions when strings of two different types collide, which has recently become important since string theory can yield cosmic strings of distinct types. Using a model containing two types of local U(1) string and stable composites, we simulate the collision of two straight strings and investigate whether the dynamics matches that previously obtained using the Nambu-Goto action, which is not strictly valid close to the junction. We find that the Nambu-Goto action performs only moderately well at predicting when the collision results in the formation of a pair of Y-junctions (with a composite string connecting them). However, we find that when they do form, the late-time dynamics matches those of the Nambu-Goto approximation very closely. We also see little radiative emission from the Y-junction system, which suggests that radiative decay due to bridge formation does not appear to be a means via which a cosmological network of such string would rapidly lose energy.
\end{abstract}

\maketitle


\section{Introduction}
Cosmic strings \cite{Vilenkin:1994, Hindmarsh:1994re} may not have played the primary role in the seeding of cosmic structure, with inflation appearing to have had that function, but they may still be important cosmological entities. Observations of, for example the cosmic microwave background (CMB) radiation, merely limit the allowed string tension. And current datasets do not do so particularly stringently: it is required that the string tension is less than about one third of that which would see them dominate the temperature anisotropies in the CMB \cite{Wyman:2005tu, Battye:2006pk, Bevis:2007gh, Battye:2007si}. Indeed they may still prove to make the primary contribution to the as-of-yet undetected CMB polarization B-mode \cite{Seljak:2006hi, Bevis:2007qz, Pogosian:2007gi}. Future CMB data, galaxy redshift surveys, gravitational wave experiments and gravitational lens surveys promise to either greatly tighten the existing constraints, or to plausibly detect cosmic strings. 

This is particularly important for (super)string/M-theory, since it has recently been realized that fundamental superstrings need not be limited to microscopic scales. Now these fundamental F-strings, along with other string theory entities called D-strings, appear able to play the role of cosmic strings \cite{Copeland:2003bj,Polchinski:2004ia}. And these cosmic superstrings would have particular properties, for example, because of the extra dimensions required by string theory, or because ($p$,$q$) bound states of $p$ F-strings and $q$ D-strings can form, with Y-shaped junctions where they unzip into two more basic constituents. As a result, the detection of cosmic strings would provide an exciting observational window upon string theory.

There has been a great deal of recent work on the topic of cosmic superstrings: from studies predicting their formation in brane inflation models \cite{Sarangi:2002yt,Jones:2003da} right through to their dynamics at late times. However, the latter case is not completely understood even for the traditional situation of gauged U(1) strings, with there being some question marks over the rate at which strings self-intersect and chop off small loops \cite{Vincent:1996rb,Vincent:1997cx,Moore:2001px,Siemens:2002dj,Martins:2005es,Ringeval:2005kr,Vanchurin:2005pa,Hindmarsh:2008}. The more complex case of cosmic superstrings is therefore particularly challenging. Various authors have used numerical simulations of field theories to represent cosmic superstrings on horizon-size scales, including the use of linear sigma models \cite{Copeland:2005cy} and global SU(2)$/\mathbb{Z}_{3}$ strings \cite{Hindmarsh:2006qn}, as well as more realistic models involving local strings \cite{Rajantie:2007hp, Urrestilla:2007yw}. These have largely addressed the question: do cosmic superstring networks evolve in the same manner as traditional gauge strings, in that their mean energy density scales with the total density of the universe?. The concern is that the bound states and Y-junctions would slow the sub-horizon decay of the strings, resulting in them dominating the universe at late times. That would, of course, be in clear contradiction with observation but fortunately the above simulations, as well as analytical modeling \cite{Tye:2005fn,Avgoustidis:2007aa}, suggest that superstring networks may exhibit scaling. However, this work is very challenging and the issue is not completely resolved. 

Given the complex array of string seen in such horizon-scale simulations, it is difficult to understand the microphysics involved in the problem and this is essential for a reliable understanding of the results. However Copeland, Kibble and Steer \cite{Copeland:2006eh, Copeland:2006if}, hereafter CKS, have recently used the Nambu-Goto approximation to study the collision between two straight strings, and have shed a great deal of light upon Y-junction formation. Unfortunately, the Nambu-Goto action assumes that the string separation and curvature scale are far greater than the string width and it is therefore not strictly valid at the site of the Y-junction itself. It cannot include, for example, the interaction between the strings, which is of course associated with the formation of the bound state. Indeed, the attraction between strings in the vicinity of the Y-junction may naively be expected to increase the bridge growth rate and to allow bridge formation when it is ruled out under the Nambu-Goto approximation. Moreover, it has been suggested \cite{Tye:2005fn} that the energy liberated by the formation of stable composites could be released as radiation and therefore help prevent the network from dominating the universe, but the Nambu-Goto action does not allow for such radiation.

Hence in this article we investigate the formation and dynamics of Y-junctions using 3D field theory simulations of a model involving two coupled Abelian Higgs models, as introduced by one of us in Ref. \cite{Saffin:2005cs}. Specifically we study the formation of bound states when two straight strings collide and then we compare our results with the analytical predictions from CKS. Reference \cite{Bettencourt:1996qe} has previously studied such collisions for a single Abelian Higgs model in the Type I regime, when bound states and Y-junctions can form, and this has been re-visited in order to test the CKS results by Ref. \cite{Salmi:2007ah}. Type I strings offer a different type of coalesce in that there is a single type of flux present, and have been studied recently for theories containing supersymmetric flat directions \cite{Cui:2007js}. Here our model contains two independent U(1) symmetries, modeling the separate F and D string charges, and hence our results are of a different nature to those of Ref. \cite{Salmi:2007ah}. Furthermore, our detailed measurements from the collision aftermath are the first to quantitatively investigate, not merely whether a composite region forms, but also its growth rate and precise dynamics. These are also important with regard to the understanding of the Y-junctions, as well as to the future application of the CKS approach.

In the next section we discuss the dynamics of Y-junctions under the Nambu-Goto approximation, before discussing the field theoretic model employed here in Sec. \ref{sec:model}. We then discuss our simulation method in Sec. \ref{sec:simulation} and our qualitative results in Sec. \ref{sec:qualitative}. Only with those results in hand can we discuss our the methods behind our detailed measurements from the simulations or the results from them, which form Secs. \ref{sec:measurements} and \ref{sec:quantitative} respectively. Finally we then present our interpretation of these results and our conclusions.


\section{Nambu-Goto dynamics}
\label{sec:NGdynamics}

\begin{figure}
\resizebox{\columnwidth}{!}{\includegraphics{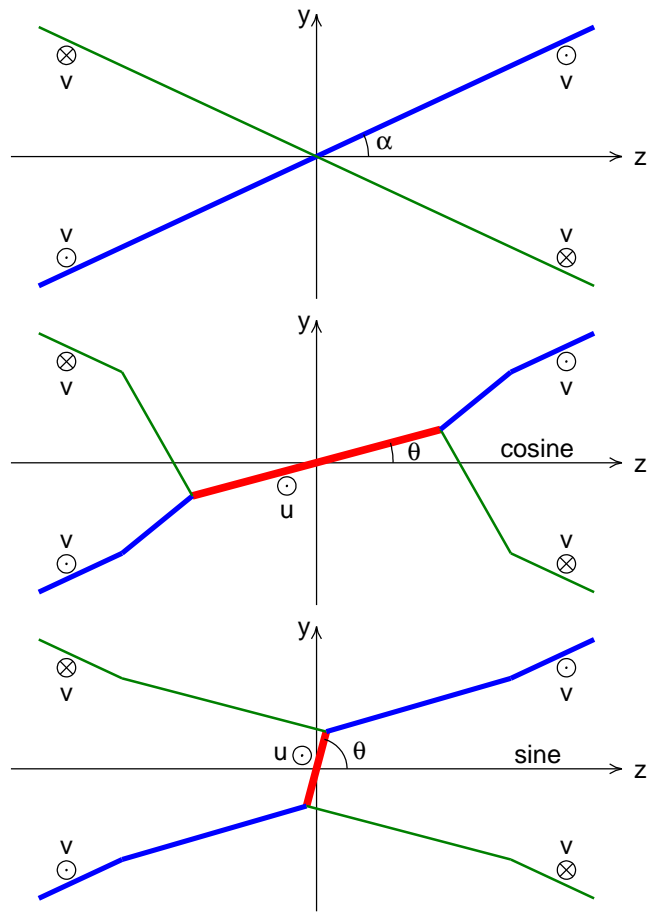}}
\caption{\label{fig:Yjunction}The intersection of two strings to form a pair of Y-junctions, as seen in the Nambu-Goto picture. The initial state of the two strings (top) is with them lying parallel to the $yz$-plane and travelling with velocities $\pm v$ in the $x$-direction, while there is a choice of final state (middle and bottom) depending upon how the two strings connect to each other. In either case, a bridge string links the two $Y$-junctions and the initial strings have kinks travelling along them. We denote the middle case as a {\it cosine} linkage, and the bottom case as a {\it sine} linkage.}
\end{figure}


The Minkowski space-time dynamics of three strings meeting at a Y-junction was solved analytically by CKS \cite{Copeland:2006eh, Copeland:2006if} under the Nambu-Goto action. Since no attraction between strings is included in this action, their approach was to add a Lagrange multiplier in order to constrain the three strings to coincide at the junction. Initial conditions may then be chosen such that there are two strings lying in the $y$-$z$ plane as shown in Fig. \ref{fig:Yjunction}(top), each making an angle $\alpha$ to the $z$-axis and travelling with velocities $v$ and $-v$ in the $x$-direction. However, the Nambu-Goto action cannot illuminate the creation of a composite bridge string upon their intersection and it must be inserted by hand how the strings connect to each other in the final state, with two such possibilities shown in Fig. \ref{fig:Yjunction}. We refer to these here as \emph{cosine} and \emph{sine} connectivities, since the essential difference between them is just the choice of placement for the angles $\alpha$ and $\theta$, and correspondingly whether cosine or sine terms appear in the primary equation below.

In general there is a third possibility for the connectivity of the strings, in which there is no partner exchange and a bridge simply grows between the two initial strings \cite{Copeland:2006eh}. However, this is forbidden in the field theoretic model that we consider here and we will not discuss it further. 

Having chosen the connectivity, the next step in the CKS method is to solve for the bridge velocity and orientation. 
If the two incident strings have energies per unit (invariant) length, $\mu_{1}$ and $\mu_{2}$, that are equal, then the symmetry present in the problem greatly simplifies matters. Firstly the two Y-junctions must remain in the $yz$-plane at all times. Secondly, the bridge must lie either along the $z$-axis (cosine link) or along the $y$-axis (sine link), and hence both the bridge speed and orientation are trivial \cite{Copeland:2006eh}.

Even in an unsymmetric case ($\mu_{1}\neq\mu_{2}$), the bridge must lie still parallel to the $yz$-plane due to the symmetry between the two Y-junctions and its velocity must be parallel to the $x$-axes \cite{Copeland:2006if}. For cosine connectivity, as shown in the middle pane of Fig \ref{fig:Yjunction}, CKS then derive the following equation for the bridge velocity $u$ \cite{Copeland:2006if}:
\begin{eqnarray}
\label{eqn:u}
0 & = & u^{4}S^{2} \sin^{2}\alpha\\ 
&& + \;u^{2}\left[ R^2 (1-v^{2}) + S^2 (v^{2}\cos^{2}\alpha - \sin^{2}\alpha)\right]\nonumber\\
&& -\;S^{2}v^{2}\cos^{2}\alpha\nonumber,
\end{eqnarray}
where we introduce the notation $R=\mu_{3}/(\mu_{1}+\mu_{2})$ and also \mbox{$S=(\mu_{1}-\mu_{2})/(\mu_{1}+\mu_{2})$}. If $\mu_{1}=\mu_{2}$, then $S$ is zero and the solution is just $u=0$, as noted above. For $S\neq0$, this equation always yields one positive root for $u^{2}$, while the sign of $u$ matches that of $S$ if string 1 initially had positive $x$-velocity. 

With $u$ having been determined, the angle $\theta$ between the bridge and the $z$-axis may then be found using \cite{Copeland:2006if}:
\begin{equation}
\label{eqn:theta}
\tan\theta = \frac{u}{v} \tan\alpha.
\end{equation}
Of course, if $S=u=0$, then $\theta$ is simply zero and, as noted above the bridge lies simply on the $z$-axis for cosine connectivity.

From these values of $u$ and $\theta$, the rate at which the length of the bridge grows may be solved for \cite{Copeland:2006if}. For $S\neq0$ it is convenient to use the invariant half-bridge length $s_{3}$ = $l_{3}/\gammaU$, where the physical bridge length is $2l_{3}$ and $\gammaU=1/\sqrt{1-u^{2}}$ is the usual Lorentz factor. This yields the neat causality constraint: $\sdot{3}<1$; since the Y-junction, which moves in the $yz$-plane at speed $dl_{3}/dt$ and at speed $u$ in the $x$-direction, cannot traverse at super-luminal speeds. A simple expression for $\sdot{3}$ was noted in Ref. \cite{Copeland:2007nv} and in our notation this is:
\begin{equation}
\label{eqn:s3dot}
\sdot{3}
=
\frac{\gammaU\cos\alpha - R\gamma \cos\theta}{\gamma\cos\theta - R\gammaU\cos\alpha}.
\end{equation}
Although there is no explicit dependence upon $S$ here, this is realized through $\gammaU$ and $\theta$, and can have a large impact on the result. However, if $S=0$ then $\gammaU$ and $\cos \theta$ are both equal to unity and may be therefore omitted, leaving a particularly compact expression \cite{Copeland:2006eh}. 

If sine connectivity is chosen instead, then it is simply the case that the replacement $\alpha\rightarrow(\pi/2-\alpha)$ and $\theta\rightarrow(\pi/2-\theta)$ should be made, or equivalently for the above equation, the cosines are merely changed to sines.

\begin{figure}
\resizebox{\columnwidth}{!}{\includegraphics{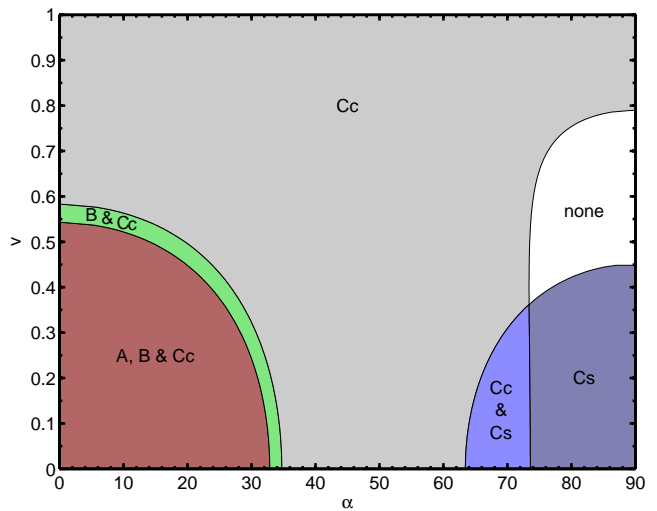}}
\caption{\label{fig:vlimits}The Nambu-Goto predictions for when cosine bridges can form in cases A, B and C that we will explore using field theory simulations, with sine bridges also included for case C. Case A has $S=0$ and $R=0.84$ while case B has $S=0.30$ and $R=0.84$. Case C has $S=0.25$ but in the field theory case the value of $R$ depends on the connectivity due to flux cancellation for a cosine bridge (Cc) with $R=0.37$ but not for a sine bridge (Cs), which then has $R=0.90$. Case Cc therefore has $|S|>R^{2}$. Note that Y-junctions under both Cc and Cs are possible outcomes for a small region of parameter space while for cases A and B there is no overlap between sine bridges (not shown) and the cosine bridges.}
\end{figure}

Of course, the solution only makes physical sense if $\sdot{3}>0$ and Fig. \ref{fig:vlimits} shows the region of the $\alpha$-$v$ plane where CKS solutions are possible for the three cases that we will explore later using field theory simulations. \mbox{Case A} has $S=0$ and $R=0.84$, and Eqn. (\ref{eqn:s3dot}) then implies that a cosine bridge can only form if $\alpha<\arccos(\gamma R)$. That is, for a given $v$ there is a certain $\alpha$ above which a CKS solution is not possible and hence it is implied that the strings must simply pass through each other. Swapping $\arccos(\gamma R)$ for $\arcsin(\gamma R)$ gives the constraint for a sine bridge, which for case A is not shown in the figure since it is just a repeat of the cosine case but with $\alpha\rightarrow(\pi/2-\alpha)$. However, if $R$ was less than $\cos 45^\circ$ then there would be a range of $\alpha$ for which growing solutions exist for both connectivities and the Nambu-Goto action would not then reveal which solution would be followed. This is something we will explore later, but only for a case with $S\neq0$.

Case B has (almost) the same value of $R$ but also has $S=0.3$. This yields, in fact, a largely similar situation with respect to the figure, although with bridge formation possible over a slightly larger area. For $|S|>R^{2}$ the situation becomes quite different and this is possible under case C. Considering firstly only cosine connectivity, the plotted case Cc highlights that CKS solutions are permitted for at high $v$ for all $\alpha$. However, a subtle point is that, in our field theory, flux cancellation (see Sec. \ref{sec:qualitative}) occurs for the cosine connectivity to yield a low $R$ of $0.37$ but this does not happen for sine connectivity and hence case Cs has a larger $R$ of $0.9$. Since the condition $|S|>R^{2}$ is not met for Cs, then this last case appears simply like cases A and B but with $\alpha\rightarrow(\pi/2-\alpha)$, although we plot it now since it is not simply a repeat of Cc. Importantly, there is a region in which both Cc and Cs solutions are possible and hence for case C we will simulate initial conditions for which the Nambu-Goto dynamics make no prediction as to which solution will actually be followed.

When bridge growth is permitted, the CKS solution for straight incident strings evolves such that the bridge grows at the constant rate specified by Eqn. (\ref{eqn:s3dot}), with the bridge orientation $\theta$ and velocity $u$ constant also. Kinks travel out along the initial strings, as shown in Fig. \ref{fig:Yjunction} and the geometry simply scales in size with the time $t$ since the collision. 


\section{U(1)$\times$U(1) dual Abelian Higgs model}
\label{sec:model}

We now explore these dynamics from a field theory perspective, including
strings which have finite width, interact strongly in the region close to the Y-junction and may radiate. To do so we employ the dual U(1) model of Ref. \cite{Saffin:2005cs}, which has also been used in horizon-volume simulations in Ref. \cite{Urrestilla:2007yw}. This involves two Abelian Higgs models coupled only via the potential term and having Lagrangian density:
\begin{eqnarray}
\label{eqn:lag}
\mathcal{L} 
& = & 
-\frac{1}{4}F_{\mu\nu}F^{\mu\nu} 
-(D_{\mu}\phi)\conj (D^{\mu}\phi)
-\frac{\lambda_{1}}{4}\left(|\phi|^{2} - \eta^{2} \right)^{2}\nonumber\\
& &
-\frac{1}{4}\mathcal{F}_{\mu\nu}\mathcal{F}^{\mu\nu} \!
-(\mathcal{D}_{\mu}\psi)\conj (\mathcal{D}^{\mu}\psi)
-\frac{\lambda_{2}}{4}\left(|\psi|^{2} - \nu^{2} \right)^{2}\nonumber\\
& &
+ \kappa \left( \left|\phi\right|^2 - \eta^{2} \right)\left( \left|\psi\right|^2 - \nu^{2} \right).
\end{eqnarray} 
We have followed the notation of Ref. \cite{Saffin:2005cs} and defined the gauge-covariant derivatives as:
\begin{eqnarray}
D_{\mu}\phi & = & \partial_{\mu}\phi - ie A_{\mu}\phi, \\ 
\mathcal{D}_{\mu}\psi & = & \partial_{\mu}\psi - ig B_{\mu}\psi, 
\end{eqnarray}
and the anti-symmetric field strength tensors as:
\begin{eqnarray}
F_{\mu\nu} & = & \partial_{\mu}A_{\nu} - \partial_{\nu}A_{\mu}, \\
\mathcal{F}_{\mu\nu} & = & \partial_{\mu}B_{\nu} - \partial_{\nu}B_{\mu}.
\end{eqnarray}
The only coupling between the two otherwise independent Abelian Higgs models is via the final potential term of Eqn. (\ref{eqn:lag}). Its form is chosen to ensure that the local U(1) symmetries associated with each Abelian Higgs model are preserved: 
\begin{eqnarray}
\phi \rightarrow \phi e^{i\omega_{\!A}}, 
& & 
A_{\mu} \rightarrow A_{\mu}+\frac{1}{e}\partial_{\mu}\omega_{\!A},
\\
\psi \rightarrow \psi e^{i\omega_{\!B}}, 
& & 
B_{\mu} \rightarrow B_{\mu}+\frac{1}{g}\partial_{\mu}\omega_{\!B},
\end{eqnarray}
since these are directly related to the presence of string solutions.

For $\kappa=0$ it is well-known that string solutions exist for each half of the model \cite{Nielsen:1973cs} (see Refs.
\cite{Vilenkin:1994, Hindmarsh:1994re} for reviews). These are characterized by the phase of $\phi$ (or $\psi$) having a net
winding of $2\pi m$ (or $2\pi n$) around any closed path that encloses the string. For $\kappa\neq0$ static and straight 
($m$,$n$) string solutions, involving both halves of the model, were found in \cite{Saffin:2005cs}. By way of an example,
consider $\nu=\eta$, $e=g$ and $\lambda_{1}=\lambda_{2}$ so that $\mu_{(0,1)}=\mu_{(1,0)}$ and also let
$2e^{2}=\lambda_{1}$ (so that the Bogomol'nyi limit \cite{Bogomolnyi:1976} applies in both halves of the model). Then if
$\kappa=0.4\sqrt{\lambda_{1}\lambda_{2}}$:
\begin{equation}
\mu_{(1,1)} = 0.840 \left[\mu_{(1,0)} + \mu_{(0,1)}\right],
\end{equation}
and hence parallel $\mu_{(0,1)}$ and $\mu_{(1,0)}$ strings can reduce their energy by combining to give a composite $\mu_{(1,1)}$ string. Smaller (positive) values of $\kappa$ yield a lower reduction in energy by composite formation, while $\kappa \geq 0.5\sqrt{\lambda_{\phi}\lambda_{\psi}}$ results in the model being unphysical, because then the potential is unbounded from below. Finally, negative $\kappa$ yields an increase in energy so composite solutions are unstable and such values are not of interest here. The values of $\mu_{(m,n)}$ for the strings involved in our simulations are shown in Table \ref{tab:mu}. 
\begin{table}
\begin{ruledtabular}
\begin{tabular}{cccc}
$m$ & $n$ & $\mu_{(m,n)} / 2\pi\eta^{2}$ & $\mu_{(m,n)} / (m+n)\mu_{(1,0)}$\\
\hline
1   &  0  & 0.864 & 1\\
1   &  1  & 1.452 & 0.840\\
2   &  0  & 1.622 & 0.938\\
2   &  1  & 2.088 & 0.805
\end{tabular}
\end{ruledtabular}
\caption{\label{tab:mu}The energy per unit length of a static string with a winding of $2\pi m$ in the phase of $\phi$ and of $2\pi n$ in the phase of $\psi$, with parameters $\lambda_{1}=\lambda_{2}=2e^{2}=2g^{2}=2$, $\eta=\nu$ and $\kappa=0.4\sqrt{\lambda_{1}\lambda_{2}}$. Note that an Abelian Higgs string of unit winding would yield $\mu=2\pi\eta^{2}$\cite{Bogomolnyi:1976}.}
\end{table}


\section{Simulation method}
\label{sec:simulation}


\subsection{Evolution algorithm}

In order to simulate the collision of two straight strings we represent the fields of the model using only their values at discrete points in space and time, and then write approximations to the second order dynamical equations in terms of the fields at these points. There is no unique way to proceed, however, for the Abelian Higgs model in Minkowksi space-time, it has become popular to use the approach of Ref. \cite{Moriarty:1988fx}. In that method, a discrete Hamiltonian is constructed and then the equations of motion for the discretized variables are obtained from it. The Hamiltonian is chosen to preserve the U(1) gauge symmetry of the Abelian Higgs model, at least in a certain discrete form. As such the simplifying gauge choice $A_{0}=0$ may be chosen but the discrete Hamilton's equation for $A_{0}$, which is the analogue of Gauss' law in the model, is preserved precisely by the remaining discrete equations. This would be difficult to achieve via the direct replacement of the derivatives in the dynamical equations with finite differences. Note also that this approach may be generalized to flat Friedman-Robertson-Walker (FRW) cosmologies via the use of a discretized action, as in Ref. \cite{Bevis:2006mj}.

Since the coupling between the two Abelian Higgs models in the present action is via the potential term only, and involves no field derivatives, then this does not greatly affect the discretization of the system. We therefore applied the above procedure to arrive at our evolution algorithm for the U(1)$\times$U(1) model in Minkowksi space-time (although our program was derived directly from the FRW code of Ref. \cite{Bevis:2006mj}). Parallel computation was made available via the use of the LATfield library \cite{LATfield}, with simulations performed across up to 32 processors of the UK National Cosmology Supercomputer \cite{Cosmos}. The (scalar\footnote{The gauge field components are represented half-way along the links between the sites ($A_{x}$ on a link parallel to the $x$-axis, $A_{y}$ on a link parallel to the $y$-axis, etc.), explicitly transporting the phase of the scalar fields across the links.}) fields were represented on the sites of a cubic lattice of spacing $\Delta x$ and therefore yielded a uniform (as opposed to adaptive) spatial resolution, while a constant timestep $\Delta t$ was also employed.


\subsection{Initial conditions}
\label{sec:ics}

We desire to start the evolution with two straight, infinite strings moving towards each other, but unfortunately there are no such analytical solutions known. Using the code written for \cite{Saffin:2005cs}, however, we can rapidly obtain numerical solutions for isolated static strings with given winding numbers, which we then employ here to construct the initial conditions following a procedure similar to that used by Ref. \cite{Moriarty:1988fx} for the Abelian Higgs model\footnote{See also Ref. \cite{Matzner:1988} for an alternative approach for the Abelian Higgs model, employing instead the Lorentz gauge, and Ref. \cite{Laguna:1990it}, which in fact employs a Lagrangian similar to Eqn. (\ref{eqn:lag}), but in a different regime in order to model superconducting strings.}.

The isolated string code solves for the radial profile of a string using a cylindrical polar coordinate system around its centre, which we denote as ($r'$,$\theta'$,$z'$) and the dashes denote that we are in the string rest frame, rather than the simulation frame. Via the appropriate choice of gauge, the profile can be written as:
\begin{eqnarray}
\phi(r',\theta',z') & = & \eta f(r') e^{im\theta'}\\
A'_{\theta}(r',\theta',z') & = & \frac{m}{e} a(r')\\
\psi'(r',\theta',z') & = & \nu p(r') e^{in\theta'}\\
B'_{\theta}(r',\theta',z') & = & \frac{n}{g} b(r'),
\end{eqnarray}
with the other components of the gauge fields simply zero. That is, for a given $m$ and $n$ this code returns the four functions $f(r')$, $a(r')$, $p(r')$, and $b(r')$ for an array of discrete $r'$ values.

In order to obtain the solution for the string in the simulation frame, that is a string moving at a speed $v$ in the $x$-direction, we must follow a similar perform a Lorentz boost. However, we also wish the string to be rotated through an angle $\alpha$ in $y$-$z$ plane, as shown in Fig. \ref{fig:Yjunction}. We therefore perform a Lorentz boost and simultaneous rotation, giving $\phi$ and $A_{\mu}$ in the simulation frame as:
\begin{eqnarray}
 \phi  & = & \phi',\\
 \label{eqn:A0}
 A_{0} & = & \gamma ( A'_{0} - v A'_{x'} ),\\
 A_{x} & = & \gamma ( A'_{x'} - v A'_{0} ),\\
 A_{y} & = & A'_{y'} \cos\alpha - A'_{z'} \sin\alpha,\\
 A_{z} & = & A'_{y'} \sin\alpha + A'_{z'} \cos\alpha.
\end{eqnarray}
Throughout this section equations for $\phi$ and $A_{\mu}$ will have obvious $\psi$ and $B_{\mu}$ analogues and we will therefore not repeatedly note that the second half of the model is to be treated in the same manner of the first. From Eqn. (\ref{eqn:A0}) it can seen that if the above rest frame solution with $A'_{0}=0$ and $r'A'_{x'}=-A'_{\theta'}\sin\theta'$ is inserted, then the resulting $A_{0}$ is non-zero. That is, the solution is not of the appropriate form for use in the $A_{0}=0$ evolution algorithm. 

Therefore, before applying the boost we perform a gauge transform $m\omega_{\!A}$ such that $A'_{x'}$ becomes zero:
\begin{equation}
\frac{1}{e} \partial_{x'} (m \omega_{A}) = - A'_{x'}.
\end{equation}
A solution to this equation is given by:
\begin{equation}
\omega_{A}(x',y') = - \frac{e}{m} \int_{0}^{x'} \!\!\! A'_{x'}(X',y',z') \; dX',
\end{equation}
and if the static solution above is inserted, this becomes:
\begin{equation}
\label{eq:aPhi}
\omega_{A}(x',y') 
= 
\int_{0}^{x'} \!\!\!  \frac{y'}{X'^{2}+y'^{2}} \; a\!\left(\sqrt{X'^{2}+y'^{2}}\right) \; dX', 
\end{equation}
which we evaluate numerically. Note that the integral is finite for all $x'$ and $y'$ since $a$ must tend to $1$ as $r'\rightarrow\infty$ in order to give finite energy per unit string length.

We choose zero as the lower integral limit because then $\omega_{A}(x',y')$ is an odd function with respect to both $x'$ and $y'$. As a result $\omega_{A}$ needs only to be found in one quadrant and can then be easily applied to all four:
\begin{eqnarray}
\label{eq:quadrant}
\omega_{\!A}(x',y')&=&-\omega_{A}(-x',y')\\
	               &=&-\omega_{A}(x',-y')\nonumber\\
                 &=&+\omega_{A}(-x',-y').\nonumber
\end{eqnarray}

Having determined $\omega_{\!A}$, then the gauge transform yields:
\begin{eqnarray}
 \phi' & = & \eta f(r') e^{im(\theta' + \omega_{\!A})}, \\
 A'_{y'} & = & \frac{m}{e} \left( \frac{1}{r'} a(r')\cos\theta' + \partial_{y'}\omega_{\!A} \right), \nonumber
\end{eqnarray}
with all other components of $A'_{\mu}$ now zero. The $y'$-derivative of $\omega_{\!A}$ can be itself written as an integral but a simpler and numerically faster procedure is to calculate $\omega_{\!A}(x',y'\pm\delta y')$ for $\delta y'$ much less than the $y'$ grid spacing and then find the gradient via a (centred) finite difference. The fields in the simulation frame are then:
\begin{eqnarray}
 \phi(x,y,z)  & = & \phi'(x',y'),\\
 A_{0}(x,y,z) & = & 0,\\
 A_{x}(x,y,z) & = & 0,\\
 A_{y}(x,y,z) & = & A'_{y'}(x',y') \cos\alpha, \\
 A_{z}(x,y,z) & = & A'_{y'}(x',y') \sin\alpha,
\end{eqnarray}
where 
\begin{eqnarray}
\label{eqn:x}
 x' & = & \gamma(x - vt),\\
 y' & = & y\cos\alpha + z\sin\alpha.
\end{eqnarray}

However, it is also required to specify the time derivatives of the fields since the equations of motion are second order. The string solution is simply translating at a velocity $v$ in the $x$-direction, therefore these derivatives can straightforwardly be obtained as:
\begin{equation}
\label{eqn:piFromDxPhi}
\partial_{t}\phi = - v \partial_{x}{\phi}
\end{equation}
where $\partial_{x}{\phi}$ can be found from $\phi(x\pm\delta x,y,z)$; and with an analogous equation for the gauge field.

Obviously, the above procedure generates only a single string and so it is required to superpose two such solutions, one with positive $v$ and one with negative $v$, but positioned such that they are initially separated by a distance much larger than the string width. Since the equations of motion are non-linear there is no precise means to do this, however for such separations a good approximation is simply to sum the single string gauge fields $A_{\mu}^{+}$ and $A_{\mu}^{-}$ to give the total:
\begin{equation}
A_{\mu} = A_{\mu}^{+} + A_{\mu}^{-}.
\end{equation}
Then for the scalar fields:
\begin{equation}
\frac{\phi}{\eta} = \frac{\phi_{+}}{\eta} \; \frac{\phi_{-}}{\eta}, 
\end{equation}
results in a superposition of complex phases and minimal interference of the core of one string due to the approximately constant, near vacuum field of the other \cite{Shellard:1987bv, Matzner:1988, Moriarty:1988fx}. Using these equations, the time derivatives must then superpose as:
\begin{eqnarray}
\partial_{t}A_{\mu} & = & \partial_{t}A_{\mu}^{+} + \partial_{t}A_{\mu}^{-},\\
\eta \; \partial_{t}\phi & = & \phi_{+}\partial_{t}\phi_{-} + \phi_{-}\partial_{t}\phi_{+}.
\end{eqnarray}
In this way two straight strings are created such that they are approaching each other and, given the form of the time dependence in Eq. \ref{eqn:x}, their centre-lines will collide at time $t=0$. That is, the initial separation is set by the choice of the simulation start time $t_{\mathrm{start}}$. 


\subsection{Boundary conditions}

Since the evolution algorithm uses finite differences to represent spatial derivatives, then the update of a field at a particular site requires the knowledge of neighbouring sites. However, at the simulation boundaries the fields are not known for all neighbours and a method must be chosen to determine a value for them. Here we simply employ our initial conditions code (but with $t \ge t_{\mathrm{start}}$) to calculate these unknown field values \cite{Matzner:1988}. Hence there is a \emph{halo} of sites surrounding the main simulation volume whose values we update using the initial conditions algorithm after each timestep.

Note that our boundary conditions will therefore reflect waves travelling along the strings and that the simulations are reliable only while the boundaries are causally unaware of the interaction between the two strings. We therefore must chose our lattice size appropriately to yield an adequate time to study the collision aftermath before artifacts of the boundary conditions reduce the reliability of the simulations.


\begin{figure}
\resizebox{\columnwidth}{!}{\includegraphics{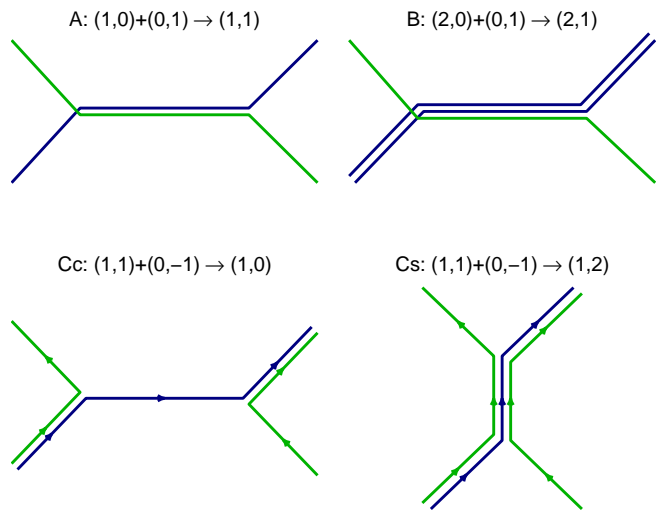}}
\caption{\label{fig:cases}Schematic illustrations of the three collision types A, B, and C considered here, with the difference between the two possible connectivities which exist for type C highlighted: Cc (cosine) and Cs (sine).}
\end{figure}

\begin{figure}
\resizebox{\columnwidth}{!}{\includegraphics{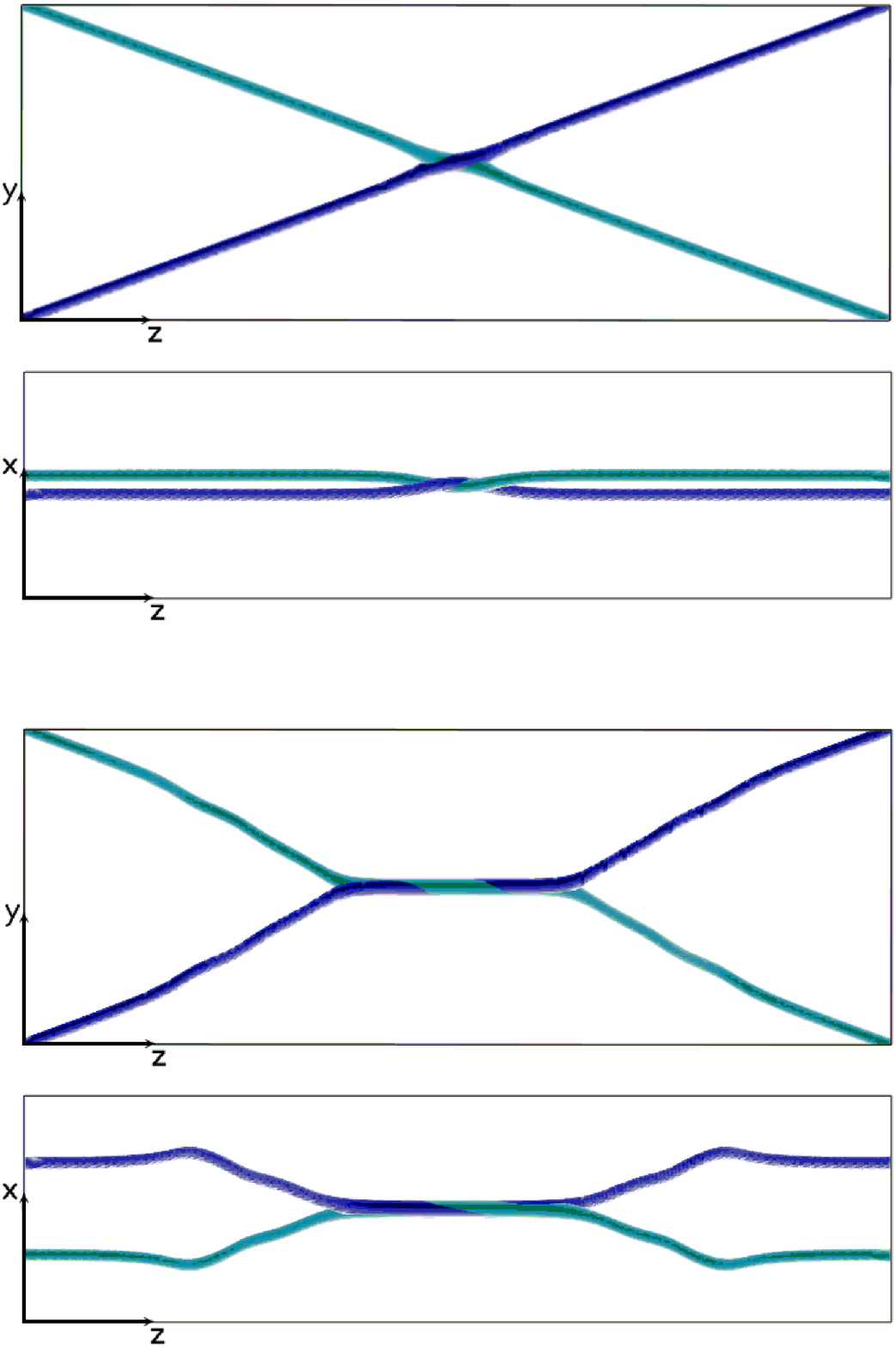}}
\caption{\label{fig:run101a} Isosurfaces of $|\phi|=0.5\eta$ and $|\psi|=0.5\nu$ from a type A collision: (1,0)+(0,1). The upper two frames are from $t=-6\eta^{-1}$, showing the attraction of the strings towards each other and intersection at negative times rather than the Nambu-Goto value of $t=0$. The lower two frames are from the later time of $t=+30\eta^{-1}$, showing the established bridge and Y-junctions. The initial speeds were $v=0.2$ and the strings made an angle $\alpha=20^{\circ}$ to the $z$-axis.}
\end{figure}

\section{Qualitative results}
\label{sec:qualitative}

Before revealing our methods for quantitative measurements of the post-collision environment we must first present our basic qualitative results. We limit ourselves to model parameters $2=\lambda_{1}=\lambda_{2}=2e^{2}=2g^{2}$ and $\kappa=0.4\sqrt{\lambda_{1}\lambda_{2}}$ and also choose equal energy scales for the two halves of our model $\eta=\nu$. As briefly mentioned in Sec. \ref{sec:NGdynamics}, we consider three different sets of initial string windings: cases A, B and C, as shown in Fig. \ref{fig:cases}. We will now discuss each one in turn.


\subsection{(1,0) + (0,1) $\rightarrow$ (1,1)?}

For the collision of a (1,0) string with a (0,1) string to plausibly yield a (1,1) string, we have $\mu_{1}=\mu_{2}$ and $R=0.840$ and we do indeed find that a composite (1,1) region forms for low values of $v$ and $\alpha$, with a Y-junction at either end and having cosine linkage. 

An example with $v=0.2$ and $\alpha=20^{\circ}$ is shown in Fig. \ref{fig:run101a}. The basic expectations from the CKS solution are indeed seen in this case, albeit with the two strings attracting each other slightly before the theoretical collision time of $t=0$, and as can be seen in the upper half of the figure, the central region of the upper string is attracted down towards the lower one (and vice-versa). The dip (or raise) breaks up into waves travelling in opposite directions\footnote{In order to understand this, it may be useful to the reader to note that $\vect{x}=\vect{x}_{+}(s+t)+\vect{x}_{-}(s-t)$ is the general solution to the equation of motion $\ddot{\vect{x}}=\partial^{2}\vect{x}/\partial s^{2}$ of Nambu-Goto strings and that for an initially static string $\vect{x}_{+}(s+t)$ is equal to $\vect{x}_{-}(s-t)$.}, resulting in low amplitude $x$-displacement patterns which move along the strings once the Y-junctions have formed. There is also a smaller disturbance travelling along each string with a displacement in the $y$-$z$ plane, created during the parallel re-alignment of the two strings at the intersection. In general such waves may be more or less pronounced and, for example, as $\alpha$ is decreased then the $t<0$ interaction has a greater $z$-range and the strings then undergo many oscillations as the strings align parallel to each other. 

At late times, when any such oscillations have subsided and the displacement waves are far from the Y-junctions, these junctions do not settle down to take on precisely the sharp Y-shape as in the Nambu-Goto case. As would be expected for interacting strings of finite width, the (1,0) and (0,1) strings curve gradually away from the $z$-axis as we move out from the junction, with the radius of curvature being a few times the string width. 

\begin{figure}
\resizebox{\columnwidth}{!}{\includegraphics{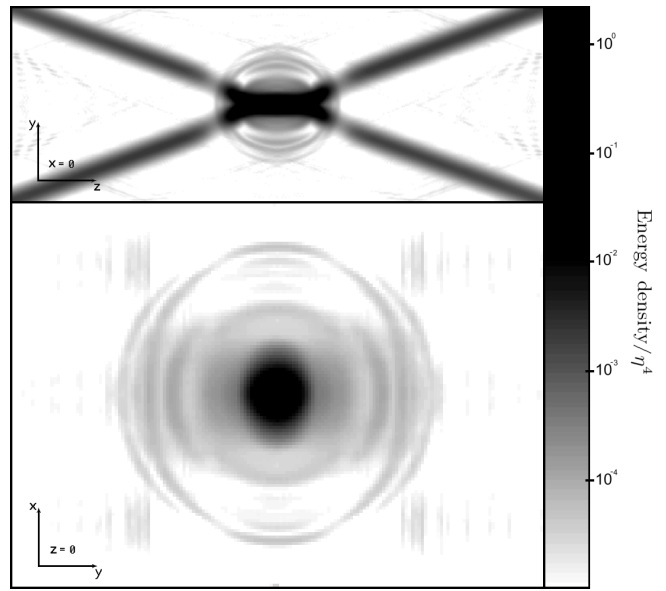}}
\caption{\label{fig:slice_201a}Energy density slices through a case A simulation, showing the initial burst of radiation as the bridge first forms. These images are heavily saturated in the bridge core, as is required to resolve the small amount of radiation emitted, and are taken from an $\alpha=20^{\circ}$, $v=0.2$ simulation at $t=21\eta^{-1}$. In the uppermost pane, the bridge appears darker than the initial strings because it lies in the $x=0$ plane, while the initial strings are $4.2\eta^{-1}$ off this plane, and note that the two panes share different spatial scales.}
\end{figure}

The formation of the bridge is accompanied by a burst of low intensity radiation, which is shown in Fig. \ref{fig:slice_201a}. Notice that the energy density in the radiation is tiny compared to that in the string core and that this represents the peak of radiation production. The initial burst soon ceases and there is then no reliably resolvable radiation produced at late times.

\begin{figure}
\resizebox{\columnwidth}{!}{\includegraphics{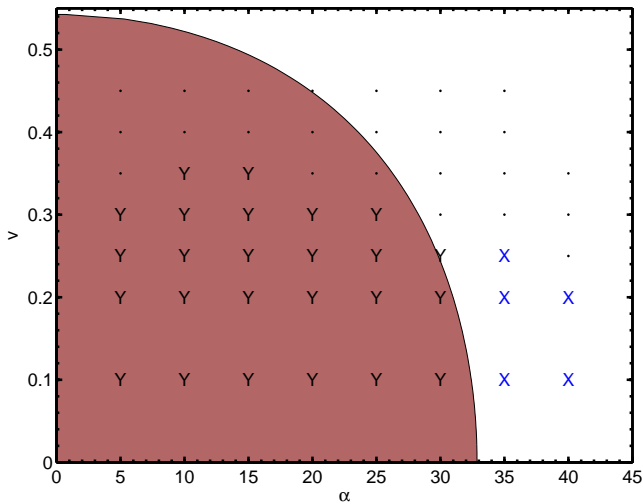}}
\caption{\label{fig:YorXcaseA}Incident velocities $v$ and angles $\alpha$ which are seen in simulations to yield Y-junctions when $\mu_{1}=\mu_{2}$ (i.e. $S=0$) and $R=0.840$. Here \textsf{Y} denotes the formation of a bridge with Y-junctions at either side, points that the strings passed through each other and \textsf{X} denotes that the strings became locked together to form an X-junction. The CKS prediction is that the region beneath the curve could yield Y-junctions.}
\end{figure}

Keeping $\alpha$ fixed and increasing $v$ reveals that, as expected from the CKS calculations, eventually the composite region no longer forms and the strings merely pass through each other. The only trace of the interaction is then displacement waves, similar to those seen when Y-junctions do form. However, the limiting $v$ for composite formation is often somewhat below the CKS prediction, as shown in Fig. \ref{fig:YorXcaseA} for this symmetric case. Note that we need only consider $\alpha<45^{\circ}$ due to the symmetry present in the initial conditions. This discrepancy between the Nambu-Goto and field theory cases is somewhat more extreme than reported by Ref. \cite{Salmi:2007ah} for the Abelian Higgs model in the type I regime, something that we will return to in our conclusions. We will also investigate the transition from bridge-forming collisions to non-bridge-forming collisions in Sec. \ref{sec:quantitative} when we discuss quantitative measurements from the simulations.

\begin{figure}
\resizebox{\columnwidth}{!}{\includegraphics{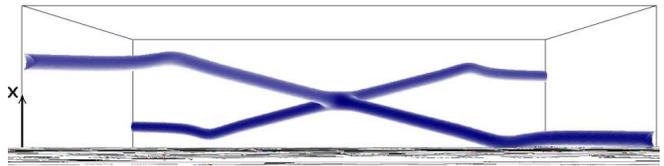}}
\caption{\label{fig:run104a}Isosurfaces of energy density $T^{0}_{0}=0.5\eta^{4}$ from a collision of type A, when $\alpha=40^{\circ}$ and $v=0.2$, showing the formation of an X-junction. Results are shown for time $t=30\eta^{-1}$.}
\end{figure}

Interestingly at larger $\alpha$ we find that there are still two possibilities for the final state of the system. For large speeds it remains the case that the strings pass through each other, while for lower values the strings become locked together and an X-junction forms, as illustrated in Fig. \ref{fig:run104a}. These are denoted by an \textsf{X} in Fig. \ref{fig:YorXcaseA}.

The Nambu-Goto solution for four strings connected in an X-junction is trivial, given the present initial conditions and the equal tensions $\mu_{1}=\mu_{2}$. The X-junction itself is simply static by symmetry and the string located between it and the kinks must then also be stationary (since the junction just reflects incident waves). However, the Nambu-Goto equations for four connected strings cannot yield any constraints because the connectivity is put in by hand. 

The field simulations show a very similar situation to these Nambu-Goto dynamics, albeit for additional minor oscillations, as in the Y-junction cases. There is only a small interaction region, but this will have a lowered energy per unit invariant length. The small amount of energy liberated by this, and a larger amount of energy liberated by the complete retardation of the incident strings, must then go into the increased length of the string, since there is again little excitation of radiative modes. However, since the binding is over only a short length of string it would be expected that even a small perturbation from these very idealized initial conditions, such as a low amplitude disturbance travelling along one of the strings, would easily break up the X-junction. The strings would then separate due to their tensions and therefore we do not believe that X-junctions would be cosmologically important in this model. They may, however, be more relevant in non-Abelian models \cite{Copeland:2006eh}.


\subsection{(2,0) + (0,1) $\rightarrow$ (2,1)?}

\begin{figure}
\resizebox{\columnwidth}{!}{\includegraphics{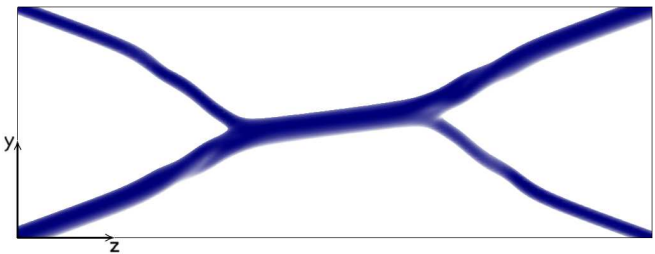}}
\caption{\label{fig:run230}Isosurfaces of energy density $T^{0}_{0}=0.5\eta^{4}$ from a type B collision: (2,0)+(0,1), with $\alpha$ and $v$ as in Fig. \ref{fig:run101a}. The snapshot is for $t=24\eta^{-1}$. Of the two initial strings, it is the (2,0) string which appears thicker and notice that the (2,1) bridge is angled toward this string.}
\end{figure}

A similar situation exists also for the unsymmetric case of a (2,0) string colliding with a (0,1) string to yield a possible (2,1) composite. Since $S=0.305$ it would be expected from CKS that, when a bridge forms, it would not be static and would not lie parallel to one of the coordinate axes but instead traverse in the same direction as the heavier initial string and be orientated closer to it. This is indeed the result apparent in Fig. \ref{fig:run230} although we return to this comparison from a quantitative perspective in Sec. \ref{sec:quantitative}.

\begin{figure}
\resizebox{\columnwidth}{!}{\includegraphics{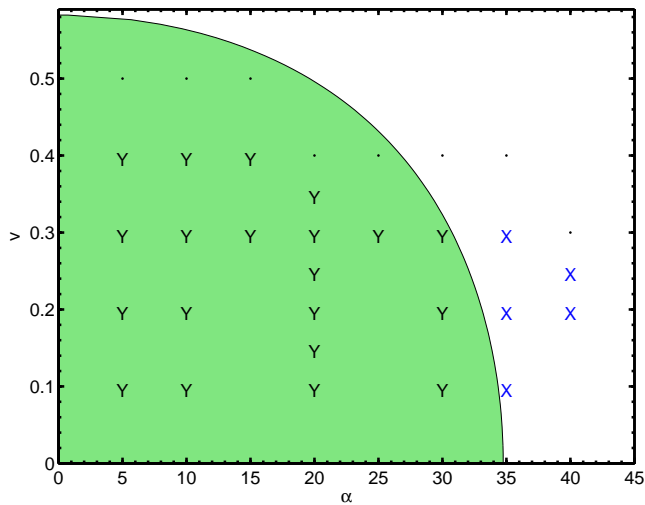}}\\
\caption{\label{fig:YorXcaseB}Incident velocities $v$ and angles $\alpha$ which are seen in simulations to yield Y-junctions when $S=0.305$ and $R=0.840$. Here \textsf{Y} denotes the formation of a bridge with Y-junctions at either side, points denote that the strings passed through each other and \textsf{X} denotes that the strings became locked together to form an X-junction. The CKS prediction is that the region beneath the curve should yield Y-junctions.}
\end{figure}

As for case A, there is a burst of radiation as the bridge forms, but again the emission appears to be weak and  limited to the bridge coalescence phase. The distribution of Y-junction formation events across the $\alpha$-$v$ plane, shown for this case in Fig. \ref{fig:YorXcaseB}, is also similar to case A. No Y-junctions are found when CKS solutions are forbidden, but the limiting velocity at a given $\alpha$ is again lower than the Nambu-Goto dynamics would allow.

Also mirroring case A, large values of $\alpha$ give X-junction formation at low $v$. Even in this unsymmetric case the Nambu-Goto solution for a X-junction is trivial and after performing a Lorentz transform to the rest frame of the junction, the strings between the junction and the kinks would become static. This appears to be be essentially what is seen in the simulations although with small oscillations as before.

\subsection{(1,1) + (0,$-1$) $\rightarrow$ (1,0) or (1,2)?}

As shown in Fig. \ref{fig:cases}, the intersection of a (1,1) string with a (0,$-1$) string can yield Y-junctions of two types depending upon the connectivity. Related to this is that the negative sign in (0,$-1$) has no absolute meaning without further qualification, and here we use the minus sign to denote that if $\alpha=0$ then the two strings carry opposing $\psi$-fluxes. Hence the naive expectation is that, for small $\alpha$, the $\psi$ fluxes will annihilate and a (1,0) bridge will form, which would yield $R=0.373$. We refer to this case as Cc, since it has cosine connectivity with respect to $\alpha$. On the other hand, for $\alpha$ close to $90^{\circ}$, the phase of $\psi$ will wind around the strings in the same direction and it might be anticipated that a (1,2) bridge will form, giving $R=0.901$. We label this case with sine linkage as Cs. For this set of initial windings, therefore, the difference between the two connectivities has an impact upon the value of $R$ and it is not simply $\cos\alpha\rightarrow\sin\alpha$ and $\cos\theta\rightarrow\sin\theta$ in Eqn. (\ref{eqn:s3dot}) for the bridge growth rate.

Since $|S|>R^2$ for a (1,0) bridge, then the Nambu-Goto prediction is that, as was shown in Fig. \ref{fig:vlimits}, bridge formation will occur for all $\alpha$ at large initial velocities ($v>0.791$). On the other hand, this is not true for the heavier (1,2) bridge and the prediction is then very much like that for cases A and B. There is, however, a small region of overlap, where both Cc and Cs allow bridge-forming CKS solutions and the Nambu-Goto equations make no predictions as to which connectivity will actually be given.

\begin{figure}
\resizebox{\columnwidth}{!}{\includegraphics{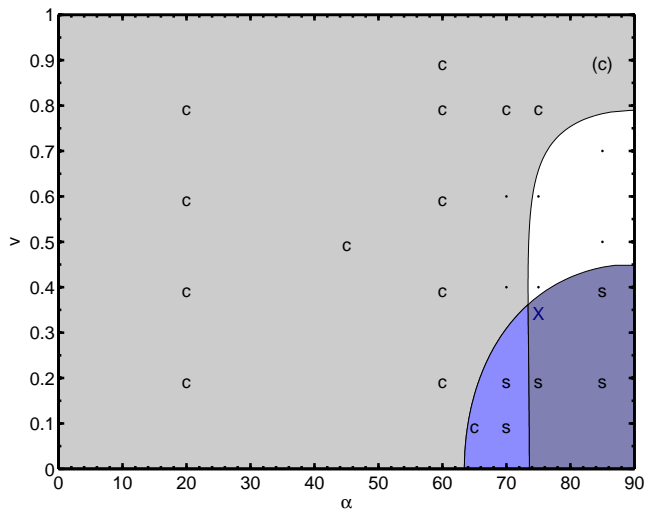}}
\caption{\label{fig:YorXcaseC}Incident velocities $v$ and angles $\alpha$ which are seen in simulations to yield Y-junctions for case C: (1,1)+(0,-1), and compared to the CKS predictions from Fig. \ref{fig:vlimits}. The final state of the system is denoted by a \textsf{c} when a (1,0) cosine bridge and Y-junction pair formed, an \textsf{s} when a (1,2) sine bridge was seen, an X when a X-junction formed and a dot when the strings passed through each other. Additionally \textsf{(c)} denotes a case when a cosine bridge formed and grew, but that a displacement wave caused a late-time intercommutation between the (1,1) and (0,-1) strings, after which the bridge collapsed.}
\end{figure}

\begin{figure}
\resizebox{\columnwidth}{!}{\includegraphics{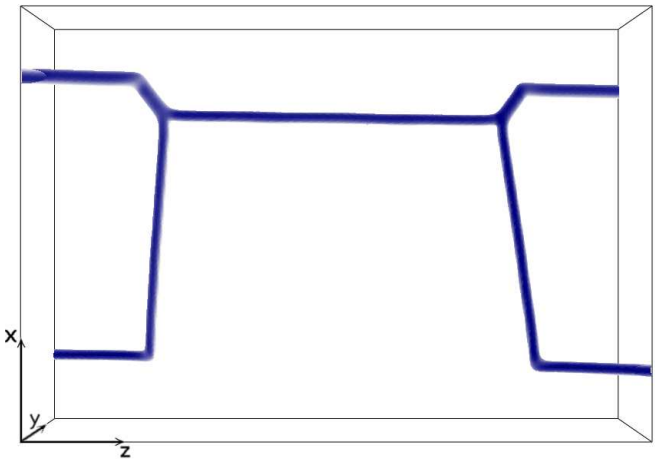}}
\caption{\label{fig:run322c}Isosurfaces of energy density $T^{0}_{0}=0.5\eta^{4}$ from the collision of (1,1) string with a (0,-1) string with initial speed $v=0.6$ and $\alpha=20^{\circ}$, showing a (1,0) bridge has formed. The snapshot is shown for $t=44\eta^{-1}$.}
\end{figure}

Indeed it is seen in the simulations that cosine bridge formation can occur at very high velocities for all values of $\alpha$ tested, as shown in Fig. \ref{fig:YorXcaseC}. An example of Cc bridge configuration is then shown in Fig. \ref{fig:run322c}. Collisions yielding sine connectivity appear, as expected, only in the bottom right of the $\alpha$-$v$ plane, while both Cs and Cc solutions are seen to populate the region in which CKS allows both solution types. There are no bridges seen in the region which CKS solutions are not possible, but as with cases A and B, the simulations show that the Nambu-Goto dynamics do not yield the boundaries between allowed and disallowed regions with total accuracy.

\begin{figure}
\resizebox{\columnwidth}{!}{\includegraphics{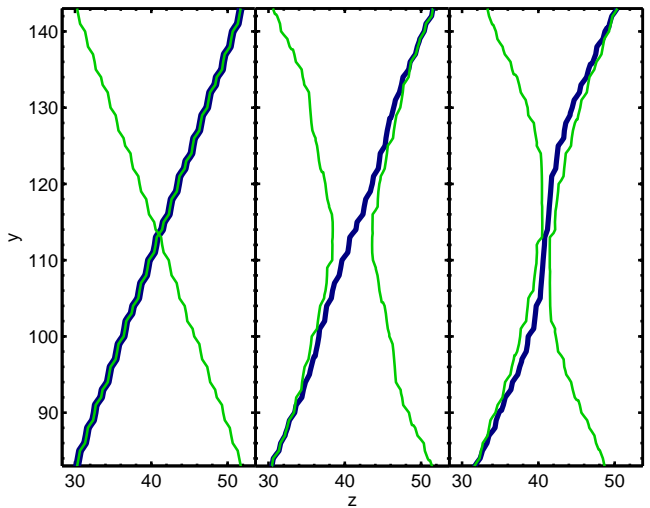}}
\caption{\label{fig:windingPath}The reconstructed string centre-lines for a type Cs collision: (1,1)+(0,-1)$\rightarrow$(1,0), occurring in the region of the $\alpha$-$v$ plane where the Nambu-Goto dynamics suggests that both Cc and Cs are possible: $\alpha=70^{\circ}$ and $v=0.2$. Results are shown for $t=-24\eta^{-1}$ (left), $t=0$ (centre) and $t=24\eta^{-1}$ (right), showing the $\phi$ string centre-line in blue and the $\psi$ centre-line in green. Centre-lines are detected using the winding of the scalar field phases around lattice plackettes (see Sec. \ref{sec:measurements} and Appendix \ref{app:interpolation}). Lengths are shown in lattice units: $\Delta x=0.5\eta^{-1}$.}
\end{figure}

The match between the CKS predictions and the simulations results is, however, somewhat closer in the present case. This is perhaps because, now differing from the first two collision types, the incident strings both have a finite $\psi$-winding. Conventional cosmic string lore would dictate that if two (0,1) strings collided, they would intercommute and therefore we have good reason to expect a significant interaction in this case, rather than the two strings passing through each another. Indeed, Fig. \ref{fig:windingPath} shows the centre-lines of the strings detected via the winding of the scalar field phases around lattice grid squares, and shows that in this type Cs collision, the first step toward the formation of a Cs bridge is an intercommutation of $\psi$-strings. The strings in the interaction region are then two (0,1) strings with a (1,0) string between them and the system has two paths via which to form Y-junctions: either (i) the $\psi$-strings upzip from the binding $\phi$-string and leave a type Cc bridge, or (ii) the $\psi$ strings zip-up further along the $\phi$ string producing a Cs bridge. The illustrated case is one in which both are possible under the Nambu-Goto approximation, but the field theory chooses the latter. Note that there is a further $\psi$-string intercommutation event before the end of the simulation but the (1,2) bridge is stable and such intercommutation events are inconsequential.

Even when neither Cc nor Cs bridges form, there is still the intercommutation between the $\psi$-strings. However, again a second intercommutation event occurs and, now in contrast with the illustrated case, the binding energy of the strings is not sufficient to hold them together so the (1,1) and (0,-1) strings simply separate.

As in collisions of type A and B, there is a burst of low intensity radiation as the bridge forms, but with the emission seeming to decay away at late times. 


\section{Measurement of the bridge length, velocity and orientation}
\label{sec:measurements}

\subsection{Symmetric case: $\mu_{1}=\mu_{2}$}

Having seen that Y-junctions form in the present model in a very similar manner to the CKS solutions, we desire to test the CKS predictions quantitatively by measuring the bridge length as a function of time, as well as its velocity and orientation. We start by detailing our method for case A since then symmetry dictates that the only possible (mean) bridge velocity is zero and the only possible (mean) orientation is parallel to either the $z$-axis or the $y$-axis. We restrict ourselves to $\alpha<\pi/4$ as before, thanks to the symmetry present in the problem, and from our previous observations need only consider bridges forming along the $z$-axis.

In order to measure the bridge length in this symmetric case we may merely count the number of sites along the line $x=y=0$ which have $|\phi|/\eta<\epsilon$ or $|\psi|/\nu<\epsilon$, with $\epsilon$ some chosen threshold.  Multiplication by $\Delta x$ and division by 2 then yields a measure of the physical \emph{half}-bridge length, which for the static bridge is equal to the invariant half-length $s_{3}$. Note that since no other string lies on this line and any initial oscillations along the bridge are only transiently important, then this is a reliable method which gives easily interpreted results.


\subsection{General case: $\mu_{1}\ne\mu_{2}$}

Unfortunately the above procedure is not readily applicable to the general case and we must therefore employ a more complex method to determine the bridge length for cases B and C, as well as additionally measuring the non-trivial bridge velocity and orientation which they yield.

In past work \cite{Vincent:1997cx, Kajantie:1998bg, Moore:2001px, Bevis:2006mj}, Abelian Higgs strings have been detected using a net winding of the scalar phase around the smallest closed loops resolved in the simulation: the lattice plackettes. However extending this to detect, for example, (1,1) strings is non-trivial since it is not guaranteed that the phase of $\phi$ has a net winding around precisely the same plackette as the phase of $\psi$. Furthermore, for (2,1) strings it is not even assured that $\phi$ will exhibit a $4\pi$ winding around a plackette rather than there being merely two close-by $2\pi$ windings. 

Our approach is therefore to firstly trace the path of every (1,0) and (0,1) elemental string through the simulation volume, such that a $4\pi$ winding in $\phi$ denotes that two (1,0) strings thread the plackette\footnote{Note that we use the gauge-invariant winding measure of Ref. \cite{Kajantie:1998bg} since in the discrete case the winding in the scalar phase can be removed by a finite gauge-transformation.}. We then detect, say a (2,1) bridge, by searching through these paths for regions in which precisely two $\phi$ string paths and one $\psi$ path string approach within a certain distance of each other.

The easiest method of reconstructing the string paths is to suppose that a straight segment of string of length $\Delta x$ passes through each plackette which has a net winding. Unfortunately constructing the paths out of an array of perpendicular segments can yield to a significant over-estimate in the total length. For the cosmological simulations of Refs. \cite{Vincent:1997cx} and \cite{Moore:2001px}, this overestimation was countered by smoothing the paths on the scale of the string width, although here we may employ a quite different approach. 

A basic observation from the simulations is that any bridge that forms is straight and, even if oscillations are present initially, these quickly decay. We therefore take the change in position vector $\Delta\vect{r}$ along each  segment and perform the vectorial sum over all detected bridge segments. We then measure the physical bridge length $2l_{3}$ as the modulus of the resulting vector, divided by the total winding ($|m|+|n|$) of the bridge:
\begin{equation}
2l_{3}=\frac{1}{|m|+|n|} \left| \sum \Delta \vect{r} \right|
\end{equation}
This effectively fits a straight line through the $|m|+|n|$ element string paths in the selected bridge region and then determines its length\footnote{Both $\phi$ and $\psi$-strings are taken to traverse the bridge in the same direction and therefore a (1,1) string, for example, would not yield erroneous cancellations in the vectorial sum.} (although note that it is in fact only a function of where the elemental paths cross the bridge detection thresholds). 

However, we additionally perform an interpolation in order to more accurately locate the intersection of the string centre-lines within the plackettes, for which we employ the method described in Appendix \ref{app:interpolation}. The vectors $\Delta \vect{r}$ are then taken to link two such intersections. Actually for our vector-based $l_{3}$ estimator this is important only for the ends of the vector and has a small effect. However, this interpolation is more important for the measurement of bridge velocity $u$ and the angle $\theta$, the methods for which are described momentarily. Furthermore, the use of our interpolation scheme and a direct $\sum |\Delta \vect{r}|$ sum of lengths yields a second $l_{3}$ estimator that we have employed as a check, and which we find performs surprisingly well.

In principle, the angle $\theta$ could be taken as that between the $z$-axis and the summed bridge vector $\sum \Delta \vect{r}$, however this would potentially yield a small systematic error. For example, where a (2,1) bridge splits into  (2,0) and (0,1) strings, the vector will be biased towards the (2,0) string and, as a result, $\theta$ will be  over-estimated. Although the effect will be small, the CKS predictions for $\theta$ show a very mild dependence upon $v$ for fixed $\alpha$, $R$ and $S$, and we wish to accurately explore this in the field theory case. We therefore perform a second vectorial sum using only the central $80\%$ of the bridge and then measure $\theta$ using:
\begin{equation}
\tan\theta = \frac{\sum_{80\%}\Delta r_{y}}{\sum_{80\%}\Delta r_{z}},
\end{equation}
that is, the angle between this new resultant vector and the $z$-axis.

Finally, in order to determine the bridge $x$-velocity $u$, we use the weighted mean of the segment $x$-coordinates over the central $80\%$ of the bridge:
\begin{equation}
\bar{x}=\frac{\sum_{80\%} (x_{+}+x_{-})|\Delta \vect{r}|}{2\sum |\Delta \vect{r}|},
\end{equation}
where $x_{\pm}$ are the endpoints of a segment. Note that the sub-plackette interpolation scheme is more important in this case since it derives from the precise values of $x_{\pm}$ even in the central regions of the bridge, rather than just at the extremes. 

Given these measurements of $l_{3}$ and $\bar{x}$, we determine $d{l}_{3}/dt$ and $u$ by performing linear fits for late times, while for $\theta$ we take the mean during the corresponding period. Note that we do not combine our measurements of $u$ and $d{l}_{3}/dt$ to yield $\sdot{3}$, but rather compare our measurements of physical length to the CKS predictions in this quantity using Eqns. (\ref{eqn:u}) and (\ref{eqn:s3dot}).


\section{Numerical Results}
\label{sec:quantitative}

We now present the results of applying the above algorithms to the three cases for which we presented qualitative results in Sec. \ref{sec:qualitative}. As before, our simulations all have $2=\lambda_{1}=\lambda_{2}=2e^{2}=2g^{2}$, $\eta=\nu$ and $\kappa=0.4\sqrt{\lambda_{1}\lambda_{2}}$.


\subsection{(1,0) + (0,1) $\rightarrow$ (1,1)}

\begin{figure}
\resizebox{\columnwidth}{!}{\includegraphics{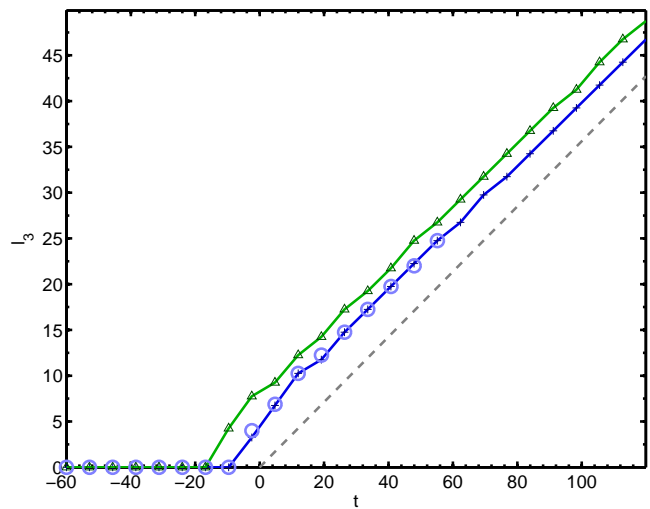}}
\caption{\label{fig:l3A}The measured $l_{3}$ as a function of $t$ measured from a $\Delta x=0.5\eta^{-1}$ simulation for a case A collision: (1,0)+(0,1), with $\alpha=20^{\circ}$ and $v=0.2$. Results shown are derived from the count of sites on \mbox{$x=y=0$} with $|\phi|<\epsilon\eta$ and $|\psi|<\epsilon\nu$ and using $\epsilon=0.5$ (blue crosses) and $\epsilon=0.75$ (green triangles), while the CKS prediction is shown by a dashed grey line. Blue cicles indicate the results for $\epsilon=0.5$ from a shorter simulation but with $\Delta x$ halved from $0.5\eta^{-1}$ to $0.25\eta^{-1}$, highlighting that the simulated dynamics are not precisely those of the continuum, while the measurements themselves are accurate to within $\Delta x/2$ and the corresponding uncertainties are too small to be shown on this plot.}
\end{figure}

In the symmetric case of a (1,0) string colliding with a (0,1) string, we find that the bridge half-length varies as shown in Fig. \ref{fig:l3A} for $\alpha=20^{\circ}$ and $v=0.2$. The CKS prediction is also indicated on the plot, being the straight line for which $l_{3}=0$ when $t=0$, and the two sets of measurements from the simulation do approximately track this. The difference between the latter two is simply due to the choice of a different threshold $\epsilon$ below which both $|\phi|/\eta$ and $|\psi|/\nu$ must be in order for a site to be considered as part of the bridge. Of course, it would be expected that the case of $\epsilon=0.75$ would show larger $l_{3}$ values than $\epsilon=0.5$ since a larger threshold will be crossed further from the string centre-lines.

If the CKS solution was precisely followed by these centre-lines, it would be expected that these two measures each would show $l_{3}=t\sdot{3}^\mathrm{CKS} + c$, where $c$ is constant and equal to the distance between the point on the $z$-axis at which the threshold is crossed and the point where the three string centre-lines meet. This is approximately what is seen, although it should be noted that the collision of the centre-lines occurs slightly before $t=0$ due to the attraction between the strings. Of course, for $t>>\eta^{-1}$ the offset $c$ will become negligible and hence from a cosmological perspective we are really only interested in whether the late-time gradient is accurately predicted by the CKS solution.

\begin{figure}
\resizebox{\columnwidth}{!}{\includegraphics{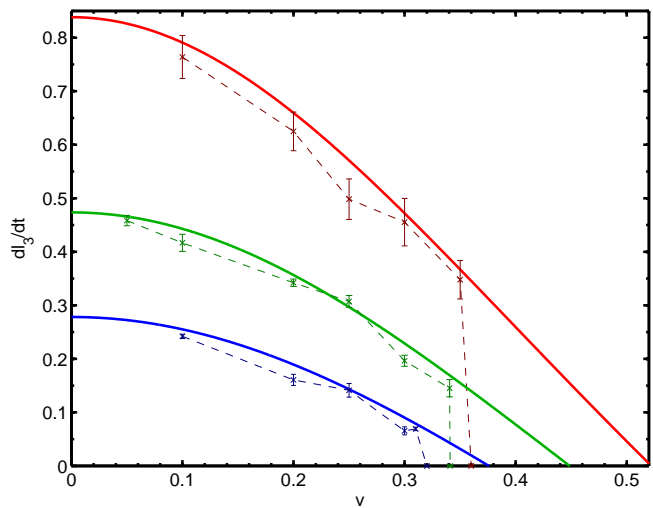}}
\caption{\label{fig:l3dotA}The measured $dl_{3}/dt$ values from a class A collision, shown as a function of $v$ for fixed $\alpha$, compared to the CKS predictions. Results are shown for $\alpha=10^{\circ}$ (uppermost), $20^{\circ}$ (middle), and $25^{\circ}$ (lower). The simulations had $\Delta x=0.5$ and were of sufficient size that signals emitted from the box centre at $t=0$ would reach the corners of the $yz$-plane by $t=60\eta^{-1}$, except for ($\alpha=20^{\circ}$, $v=0.2$), ($\alpha=20^{\circ}$, $v=0.3405$) and ($\alpha=25^{\circ}$, $v=0.31$) which ran till $t=120$, $90$ and $120\eta^{-1}$.}
\end{figure}

The measured gradients for various $\alpha$ and $v$ values are shown in Fig. \ref{fig:l3dotA} and compared to the CKS predictions. An approximate uncertainty estimation is performed such that $dl_{3}/dt$ is taken from a linear fit to the final third of the apparent linear region, with the error bar shown being the standard deviation across the three thirds. This method is sensitive to both systematic differences between the early- and late-time dynamics and to the measurement uncertainties. The plot shows that when Y-junctions form there is excellent agreement between the simulations and the Nambu-Goto predictions.  However if $v$ is increased at fixed $\alpha$, then there is a certain critical value $v_{\mathrm{c}}$ where $dl_{3}/dt$ drops suddenly to zero and away from the CKS solution. There is not, as one might have expected, a gradual divergence from the CKS predictions, and the results seen in Fig. \ref{fig:YorXcaseA} are not due to $dl_{3}/dt$ in our model slowly falling away from the Nambu-Goto value and reaching $dl_{3}/dt=0$ at a lower value of $v$. Instead it is the case that: either Y-junctions do not form at all, or they form and an $d{l}_{3}/dt$ value equal to (or at least extremely close to) the CKS value is observed.


\subsection{(2,0)+(0,1) $\rightarrow$ (2,1)}

\begin{figure}
\resizebox{\columnwidth}{!}{\includegraphics{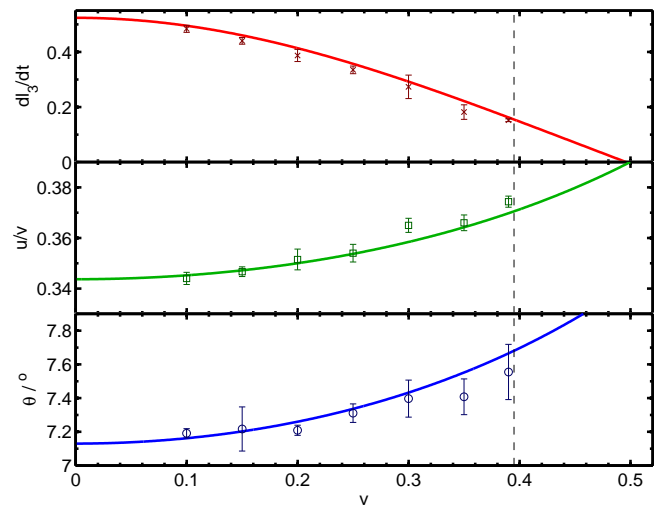}}
\caption{\label{fig:l3dotUthetaB}The measured physical bridge half-length growth rate $dl_{3}/dt$, bridge speed $u$ and bridge orientation $\theta$ from type B collisions with $\alpha=20^{\circ}$, compared to CKS predictions. The simulation had $\Delta x=0.5$ and were of sufficient size that signals emitted from the box centre at $t=0$ would reach the corners of the $yz$-plane by $t=60\eta^{-1}$ (except for $v=0.39$ when this value was $120\eta^{-1}$).}
\end{figure}

The measurements of $dl_{3}/dt$, $u$ and $\theta$ from a collision of type B are shown as a function of $v$ for fixed $\alpha=20^{\circ}$ in Fig. \ref{fig:l3dotUthetaB}, with uncertainties estimated via the same approximate method as for case A. Results indicate that, as in the type A collisions, if Y-junctions form then these three measures are well-predicted by the Nambu-Goto physics. Even the slight departure from $u\propto v$, a case which would appear on our $u/v$ plot as a horizontal line, is quite obvious and in concordance. However, resolving the change in $\theta$ with $v$, at fixed $\alpha$, is more challenging, being just $0.6^{\circ}$ across the entire range over which Y-junctions are seen to form. Hence our simulations have struggled to concretely resolve the trend, although they do give an indication of the slight increase at  larger $v$. It did not seem worthwhile to utilize very large simulations in order to reduce the measurement uncertainties since it is surely of little consequence whether the CKS predictions get the bridge orientation wrong by small fractions of a degree, and in any case, we explore this variation more completely for case C.

As in case A, the plot shows that when $v$ is increased to a certain critical value, the CKS predictions suddenly fail and Y-junctions do not form for greater speeds. Again there is no gradual reduction in the late-time $d{l}_{3}/dt$ value. However in this case, with symmetry not fixing $u$ and $\theta$, we see additionally that these two appear to match the CKS predictions right up to this critical speed. 

\subsection{(1,1)+(0,-1) $\rightarrow$ (1,0)}

\begin{figure}
\resizebox{\columnwidth}{!}{\includegraphics{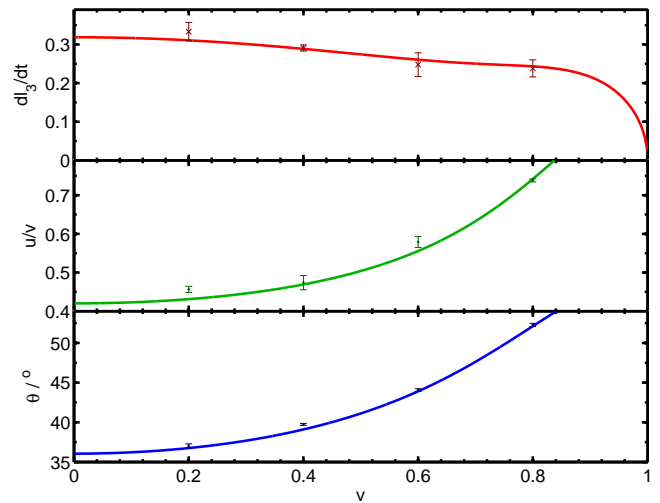}}
\caption{\label{fig:l3dotUthetaC}The measured physical bridge half-length growth rate $dl_{3}/dt$, bridge speed $u$ (blue) and bridge orientation $\theta$ (red) values from type Cc collisions with $\alpha=60^{\circ}$, compared to CKS predictions. The simulations at low speed had $\Delta x=0.5$, which was then decreased to accommodate Lorentz contraction, with the simulation size such that signals emitted from the box centre at $t=0$ would reach the corners of the $yz$-plane by $t=60\eta^{-1}$.}
\end{figure}

Case Cc provides a quite different collision, in that $|S|>R^{2}$, and also the initial strings both have a net winding in $\psi$. As shown in Fig. \ref{fig:l3dotUthetaC}, the $dl_{3}/dt$ results are for type Cc collisions at $\alpha=60^{\circ}$ are in complete concordance with the CKS predictions. This agreement appears to be maintained also for both $u$ and $\theta$, with the more sensitive dependence of the latter upon $v$, relative to case B, clearly resolved by these data.

\section{Discussion of the results}

As noted earlier, the Nambu-Goto action is not valid in the vicinity of a Y-junction and ignores, for example, the attraction between the strings in that region. In this section, however, we present a discussion of why we would in fact expect the Nambu-Goto dynamics to be a good description of the late-time behaviour seen in our simulations. To do so we first re-derive the CKS solution using energy and momentum conservation for Nambu-Goto strings, rather than the equations of motion, and then we discuss the changes that the field theory case would yield. A corollary of this will be a direct physical interpretation of the CKS constraints on Y-junction formation from the intersection of strings.

\subsection{Energy-momentum derivation of the CKS solution}
\label{subsec:tensions}

In order to derive the CKS solution from energy and momentum conservation, we must first consider the energy-momentum tensor $T^{\mu}_{\nu}$ of a Nambu-Goto string with energy per invariant unit length $\mu$. Labelling space-time coordinates as $(t,x,y,z)$ and considering the string to lie along the $z$-axis while travelling at speed $v$ in the $x$-direction, then:
\begin{equation}
\label{eqn:T}
T^{\mu}_{\nu}
=
\mu
\left(\begin{array}{cccc}
\gamma  & -v\gamma    & 0 & 0 \\
v\gamma & v^{2}\gamma & 0 & 0 \\
0       & 0           & 0 & 0 \\
0       & 0           & 0 & 1/\gamma
\end{array}\right)
\delta(x)
\delta(y).
\end{equation}
That is, the $x$-momentum per unit physical length is $\mu v\gamma$ and so per unit invariant it is just $\mu v$.
The term $T^{z}_{z}$ is the tension of the string, and it is important to note that this is reduced by a factor $\gamma$ for a moving string and is not in general equal to the mass per unit invariant length.

\begin{figure}
\resizebox{\columnwidth}{!}{\includegraphics{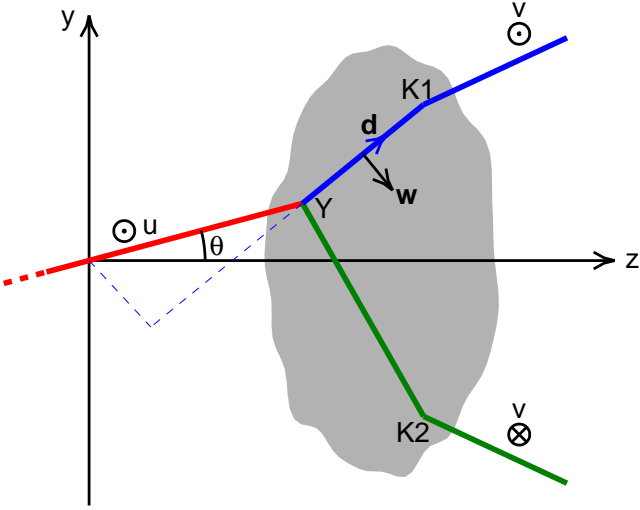}}
\caption{\label{fig:geometry}The geometry involved in the tension-based interpretation of Eqn. (\ref{eqn:s3dot}). The Y-junction is at position Y and the two kinks are at K1 and K2, while the region upon which the tensions in question act is highlighted in gray. The dashed blue lines indicate the consumption of the string 1 due to the motion of Y. (Note that both the unit vector $\vect{d}$ along the line YK1 and the orthogonal velocity $\vect{w}$, have in general finite components in the $x$-direction, therefore the projections of them onto the $y$-$z$ plane would not normally be perpendicular, although we shown them as orthogonal for illustrative purposes)}
\end{figure}

We will allow some guiding input from the CKS solution and assume that the system has the geometry shown in Fig. \ref{fig:geometry}, with straight strings except for kinks at K1 and K2. We therefore write the equation for the line YK1 as: 
\begin{equation}
\vect{r}=\vect{w}t+\frac{\sigma}{\gammaW} \vect{d},
\end{equation}
where $\vect{w}$ is the velocity of the string, $\vect{d}$ is a unit vector pointing from Y to K1 and $\sigma$ is the invariant length measured from the point $\vect{w}t$. Note that we take $\vect{w}\cdot\vect{d}=0$, which is effectively a choice of gauge. 

The Y-junction lies at position:
\begin{equation}
\label{eqn:rY}
\vect{r}_{\mathrm{Y}} 
= 
\vect{w}t + \frac{\sY}{\gammaW} \vect{d}
=
\left( \begin{array}{c} 
u \\ \sdot{3} \sin\theta / \gammaU \\ \sdot{3} \cos\theta / \gammaU
\end{array} \right) t.
\end{equation}
If Y moves such that $\sY$ becomes positive, then invariant length is removed from string 1. Following the notation of CKS we label the \emph{production} rate of string 1 at the Y-junction as $\sdot{1}$ and hence $\sY=-\sdot{1}t$. We also have that K1 travels at the speed of light and therefore its position is: 
\begin{equation}
\label{eqn:rK1}
\vect{r}_{\mathrm{K1}}
=
\vect{w}t+\frac{\sK}{\gammaW} \vect{d}
=
\left( \begin{array}{c} 
v \\ \sin\alpha / \gamma \\ \cos\alpha / \gamma
\end{array} \right) t.
\end{equation}
The value of $\sK$ is given by the invariant string length traversed by the kink per unit time and is simply $\sK=t$. Therefore the invariant length of the line YK1 is just $(\sK-\sY)=(1+\sdot{1})t$.

In principle, from these line end-points, we can now find expressions for $\sdot{1}$, $\vect{w}$ and $\vect{d}$ in terms of $\sdot{3}$, $u$, and $\theta$; while applying a corresponding procedure for the line YK2. However, for the present discussion we only need to determine the total $z$-momentum from these two lines. For the first of them this is revealed by the addition of the last two equations as $\mu_{1}$(\ref{eqn:rY})+$\mu_{1}\sdot{1}$(\ref{eqn:rK1}), resulting in:
\begin{equation} 
p_{z}^{\mathrm{YK1}}
=
\mu_{1}(1+\sdot{1})t w_{z} 
= 
\mu_{1}\sdot{3}\frac{\cos\theta}{\gammaU}t + \mu_{1}\sdot{1}\frac{\cos\alpha}{\gamma}t
\end{equation}
Hence the total $z$-momentum from both YK1 and YK2 is:
\begin{equation}
p_{z}
=
(\mu_{1}+\mu_{2}) \frac{\sdot{3}\cos\theta}{\gammaU} t
+ 
(\mu_{1}\sdot{1}+\mu_{2}\sdot{2}) \frac{\cos\alpha}{\gamma} t.
\end{equation}
This can be simplified by noting that conservation of energy at Y implies that:
\begin{equation}
\label{eqn:consEnergy}
\mu_{1}\sdot{1}+\mu_{2}\sdot{2}+\mu_{3}\sdot{3} = 0
\end{equation}
in the Nambu-Goto case, since there is no radiative emission. Therefore we can write the combination of $\sdot{1}$ and $\sdot{2}$ in terms of $\sdot{3}$, yielding the rate of change of $z$-momentum as:
\begin{equation}
\label{eqn:dotpz}
\dot{p}_{z}
=
\sdot{3}\left(
(\mu_{1}+\mu_{2})\frac{\cos\theta}{\gammaU} 
-
\mu_{3}\frac{\cos\alpha}{\gamma}
\right).
\end{equation}
That is, the growth of the bridge necessarily leads to an accumulation of $z$-momentum along these two lines. 

Since the incident strings in the regions beyond the kinks simply continue with velocities $\pm v$ in the $x$-direction, and symmetry requires the bridge to travel in the $x$-direction also, then the only $z$-momentum in the highlighted region of Fig. \ref{fig:geometry} is that due to the strings along YK1 and YK2. This is provided by the $z$-components of the tensions that act externally on the highlighted region:
\begin{equation}
F_{z} 
= 
(\mu_{1}+\mu_{2}) \frac{\cos\alpha}{\gamma} - \mu_{3} \frac{\cos\theta}{\gammaU},
\end{equation}
and therefore we immediately have a physical condition for bridge growth, simply:
\begin{equation}
F_{z} > 0.
\end{equation}
Note that this is not the sum of tensions at the Y-junction itself, which might naively be expected to be the relevant tensions for the growth condition. It is also not necessarily true for the tension components along the direction in which bridge growth actually occurs.

Of course, if we now set $F_{z}$ equal to the rate of change of momentum, then we trivially obtain Eqn. (\ref{eqn:s3dot}). That is, we may derive the expression for $\sdot{3}$ based upon the assumptions that the string energy and momentum are conserved, straight strings join the Y-junction to the regions causally disconnected from it, the bridge is straight\footnote{Note that if the bridge is straight for all time then it must have a velocity that is uniform along its length and the velocity must then be in the $x$-direction by symmetry, hence this is not an independent assumption.}, and that the energy-momentum tensor is that of Eqn. (\ref{eqn:T}).

\subsection{Explanation of the accuracy of the Nambu-Goto approximation}

The above assumptions will not be precisely met in the field theory case, but now suppose that, even then, the geometry of the system at a certain time $t$ is approximately as shown in Fig. \ref{fig:geometry}. However, allow for differences relative to the Nambu-Goto case including that the kinks at K1 and K2 are smoothed on a scale close to the string width and that also there are displacement waves in their vicinity, which were left over from the initial bridge formation. Suppose additionally that there will be a significant attraction between the strings in the region close to the Y-junction and that their paths will gradually curve towards each other rather than there being a sharp Y-shape. 

So long as $t>>\eta^{-1}$, these changes will have little effect on the large-scale geometry. Therefore we may start to follow the above discussion. We must, however, now ask ourselves whether energy-momentum conservation at Y would take the same form as Eqn. (\ref{eqn:consEnergy}). Even with the interaction region close to Y, if the shape of the strings remains constant in time and simply translates with speed $d{l}_{3}/dt$, then the energy associated with the interaction would not change. Additionally, if there is merely translation of this region, there is no excitation of radiation. Hence it appears that the energy conservation equation \emph{could} be unchanged.

Therefore, while the $z$-momentum in the shaded region of Fig. \ref{fig:geometry} would not follow precisely Eqn. (\ref{eqn:dotpz}), for late times this difference might be negligible. Further, the tensions external to the shaded region would be simply the same as in the Nambu-Goto case (for a particular $\theta$ and $u$) and hence equating their $z$-components with the rate of change of $z$-momentum, we then would obtain $\sdot{3}$ as being very close to Eqn. (\ref{eqn:s3dot}). Therefore we expect that a plausible late-time solution for the field theory case is one which tends towards a CKS solution, despite the interaction near Y. That is, of course, assuming that a physical CKS solution with $\sdot{3}>0$ exists for the given $\alpha$ and $v$.

However, these arguments do not guarantee that the Nambu-Goto dynamics will be a good description of the system and the initial distortion of the strings due to their attraction and the associated oscillations mean that the strings do not initially follow the above geometry all that closely. There is also an initial burst of radiation, as well as the finite width of the strings, both of which complicate the field theoretic case. As we have seen, the CKS solution appears to be only a reliable indicator of when Y-junction formation is \emph{not} possible, and of the late-time dynamics when it is.


\section{Conclusions}

Our results indicate that for the present field theory, involving two coupled Abelian Higgs models, the CKS solutions for Nambu-Goto strings give largely accurate predictions as to whether or not Y-junctions will form when two straight strings collide. Like the results found in Ref. \cite{Salmi:2007ah} for the Abelian Higgs model in the type I regime, we find that they are not entirely accurate, however, for two of the three initial string pairings studied here we see a noticeably poorer match than in that reference. We believe that this difference is due to the fact that for our third collision type, both incident strings had finite $\psi$-winding and therefore an initial intercommutation interaction was highly likely, even with the presence of the $\phi$-half of the model. This initial interaction appears to aid the alignment of the strings and the initial formation of the Y-junction. This observation is further compounded by noting that in the Abelian Higgs model studied by Ref. \cite{Salmi:2007ah}, the bridge formation is also preceded by an intercommutation event. Hence, we note that the applicability of the CKS bridge formation predictions to field theory models is likely to be sensitive to the model employed, and as we have seen, to the exact nature of the strings involved. 

However, in all three of our collision types we find that when Y-junctions did form, the late-time dynamics of the system was very accurately described by the corresponding CKS solution. Further, given the discussion in the previous section, we believe that the model dependence of the initial bridge formation is likely to be less significant for the late-time dynamics. Therefore, despite the break-down of the Nambu-Goto approximation near the Y-junction, the CKS approach appears to be a very powerful method for studying Y-junctions in local string models. On the other hand, global strings, with significant long range interactions are likely to be poorly represented by the Nambu-Goto dynamics.

We find that, in our model, bridge formation is accompanied by a short period of weak radiative emission, but find little evidence of such emission at late times. Hence this suggests that radiative decay due to bridge creation is unlikely to be an effective means of dissipating the energy in a cosmological string network. Our results therefore offer little comfort to authors who fear that networks of string capable of bridge formation might ``freeze out'' and, in contradiction with observation, grow to dominate the universe. While our radiative emission conclusions are only relevant for the model studied here, our arguments from the previous section also suggest that local strings in general will mirror the results here.

A Nambu-Goto simulation using the CKS method might well assist the simulation of bridge-forming models over horizon scales, since in principle it would avail a greater dynamic range than is possible in field theory simulations. This is more relevant for these superstring-inspired models than for traditional U(1) cosmic strings, since the computational outlay for the additional fields present is greater, but also the complex network dynamics require study over a longer time period. However with any such simulations, great care must be taken to note the differences in the results between Nambu-Goto simulations for traditional strings and the U(1) field theory counterparts, which automatically include a greater depth of physics \cite{Vincent:1997cx, Moore:2001px, Hindmarsh:2008}. These differences are, however, largely at small scales and do not preclude the usefulness of Nambu-Goto simulations with Y-junctions. The long term aim must be, of course, to link the string network properties to observations and to assess the difference between the signatures of traditional U(1) strings and cosmic superstrings.


\begin{acknowledgments}
We acknowledge financial support from STFC (both N.B. and P.S.). Production simulations were run on the UK National Cosmology Supercomputer, supported by SGI, Intel, HEFCE and STFC. We thank Ed Copeland and Jon Urrestilla for useful discussions.
\end{acknowledgments}

\appendix

\section{Simple sub-plackette interpolation method}
\label{app:interpolation}

Given a net winding of the phase of $\phi$ around a lattice plackette, there is no unique means of using the magnitudes and phases of the corresponding scalar field at the four plackette corners in order to determine to location of $\phi=0$ on the plackette surface. In principle one may use the known string profile to perform a best-fit determination of the position and orientation of the string centre-line as it intersects the plackette, but for present purposes we do not require such a complex procedure. Here we use a simple and computationally rapid method that, importantly, is guaranteed to yield $\phi=0$ coordinates that lie within the plackette grid square. That is, we estimate the coordinates as:
\begin{equation}
\bar{\vect{x}} = \frac{\sum_{i=1}^{4} \vect{x}_{i} |\phi_{i}|^{-1/m}}{\sum_{i=1}^{4} |\phi_{i}|^{-1/m}},
\end{equation}
where $\vect{x}_{i}$ are the positions of the four plackette corners, $\phi_{i}$ the field values at them, and $2\pi m$ is the net winding around the plackette. The index $-1/m$ is included in this expression because, close to its axis, an ideal static \mbox{($m$, $n$)} string has:
\begin{equation}
|\phi_{i}| \propto r^{m},
\end{equation}
where $r$ is the radial coordinate. This is exactly as in the Abelian Higgs model, even though there is the additional coupling in the present case. The $\bar{\vect{x}}$ expression above therefore involves an approximate $r^{-1}$ weighting of the plackette corners so that $\bar{\vect{x}}$ will be drawn towards to those which appear to lie closer to the string centre-line. 

A clear downside to this method is that, for a plackette with sides in the $x$ and $y$ directions, the value of $\bar{x}$ varies with the true $y$-coordinate of the $\phi=0$ point, however, it is essentially a zero-cost means of improving the string paths through the simulation volume. Unlike a smoothing operation, this process encapsulates greater information from the fields in the simulation and, for the example of a straight string of constant $x$-coordinate, improves our knowledge of that $x$-coordinate, which smoothing cannot achieve. It is certainly sufficient for its application in the present article.

A by-eye assessment of its usefulness is afforded by Fig. \ref{fig:windingPath}, in which a series of $90^{\circ}$ (or $45^{\circ}$) steps is not seen and instead the centre-line paths are relatively smooth. It is of course the case, however, that some obtuse kinks are visible.


\bibliography{YjunctionSimulations}

\begin{thebibliography}{44}
\expandafter\ifx\csname natexlab\endcsname\relax\def\natexlab#1{#1}\fi
\expandafter\ifx\csname bibnamefont\endcsname\relax
  \def\bibnamefont#1{#1}\fi
\expandafter\ifx\csname bibfnamefont\endcsname\relax
  \def\bibfnamefont#1{#1}\fi
\expandafter\ifx\csname citenamefont\endcsname\relax
  \def\citenamefont#1{#1}\fi
\expandafter\ifx\csname url\endcsname\relax
  \def\url#1{\texttt{#1}}\fi
\expandafter\ifx\csname urlprefix\endcsname\relax\def\urlprefix{URL }\fi
\providecommand{\bibinfo}[2]{#2}
\providecommand{\eprint}[2][]{\url{#2}}

\bibitem[{\citenamefont{Vilenkin and Shellard}(1994)}]{Vilenkin:1994}
\bibinfo{author}{\bibfnamefont{A.}~\bibnamefont{Vilenkin}} \bibnamefont{and}
  \bibinfo{author}{\bibfnamefont{E.~P.~S.} \bibnamefont{Shellard}},
  \emph{\bibinfo{title}{Cosmic Strings and Other Topological Defects}}
  (\bibinfo{publisher}{Cambridge University Press, Cambridge, U.K.},
  \bibinfo{year}{1994}).

\bibitem[{\citenamefont{Hindmarsh and Kibble}(1995)}]{Hindmarsh:1994re}
\bibinfo{author}{\bibfnamefont{M.~B.} \bibnamefont{Hindmarsh}}
  \bibnamefont{and} \bibinfo{author}{\bibfnamefont{T.~W.~B.}
  \bibnamefont{Kibble}}, \bibinfo{journal}{Rept. Prog. Phys.}
  \textbf{\bibinfo{volume}{58}}, \bibinfo{pages}{477} (\bibinfo{year}{1995}),
  \eprint{hep-ph/9411342}.

\bibitem[{\citenamefont{Wyman et~al.}(2005)\citenamefont{Wyman, Pogosian, and
  Wasserman}}]{Wyman:2005tu}
\bibinfo{author}{\bibfnamefont{M.}~\bibnamefont{Wyman}},
  \bibinfo{author}{\bibfnamefont{L.}~\bibnamefont{Pogosian}}, \bibnamefont{and}
  \bibinfo{author}{\bibfnamefont{I.}~\bibnamefont{Wasserman}},
  \bibinfo{journal}{Phys. Rev.} \textbf{\bibinfo{volume}{D72}},
  \bibinfo{pages}{023513} (\bibinfo{year}{2005}), \eprint{astro-ph/0503364}.

\bibitem[{\citenamefont{Battye et~al.}(2006)\citenamefont{Battye, Garbrecht,
  and Moss}}]{Battye:2006pk}
\bibinfo{author}{\bibfnamefont{R.~A.} \bibnamefont{Battye}},
  \bibinfo{author}{\bibfnamefont{B.}~\bibnamefont{Garbrecht}},
  \bibnamefont{and} \bibinfo{author}{\bibfnamefont{A.}~\bibnamefont{Moss}},
  \bibinfo{journal}{JCAP} \textbf{\bibinfo{volume}{0609}}, \bibinfo{pages}{007}
  (\bibinfo{year}{2006}), \eprint{astro-ph/0607339}.

\bibitem[{\citenamefont{Bevis et~al.}(2008)\citenamefont{Bevis, Hindmarsh,
  Kunz, and Urrestilla}}]{Bevis:2007gh}
\bibinfo{author}{\bibfnamefont{N.}~\bibnamefont{Bevis}},
  \bibinfo{author}{\bibfnamefont{M.}~\bibnamefont{Hindmarsh}},
  \bibinfo{author}{\bibfnamefont{M.}~\bibnamefont{Kunz}}, \bibnamefont{and}
  \bibinfo{author}{\bibfnamefont{J.}~\bibnamefont{Urrestilla}},
  \bibinfo{journal}{Phys. Rev. Lett.} \textbf{\bibinfo{volume}{100}},
  \bibinfo{pages}{021301} (\bibinfo{year}{2008}), \eprint{astro-ph/0702223}.

\bibitem[{\citenamefont{Battye et~al.}(2007)\citenamefont{Battye, Garbrecht,
  Moss, and Stoica}}]{Battye:2007si}
\bibinfo{author}{\bibfnamefont{R.~A.} \bibnamefont{Battye}},
  \bibinfo{author}{\bibfnamefont{B.}~\bibnamefont{Garbrecht}},
  \bibinfo{author}{\bibfnamefont{A.}~\bibnamefont{Moss}}, \bibnamefont{and}
  \bibinfo{author}{\bibfnamefont{H.}~\bibnamefont{Stoica}}
  (\bibinfo{year}{2007}), \eprint{arXiv:0710.1541 [astro-ph]}.

\bibitem[{\citenamefont{Seljak and Slosar}(2006)}]{Seljak:2006hi}
\bibinfo{author}{\bibfnamefont{U.}~\bibnamefont{Seljak}} \bibnamefont{and}
  \bibinfo{author}{\bibfnamefont{A.}~\bibnamefont{Slosar}},
  \bibinfo{journal}{Phys. Rev.} \textbf{\bibinfo{volume}{D74}},
  \bibinfo{pages}{063523} (\bibinfo{year}{2006}), \eprint{astro-ph/0604143}.

\bibitem[{\citenamefont{Bevis et~al.}(2007{\natexlab{a}})\citenamefont{Bevis,
  Hindmarsh, Kunz, and Urrestilla}}]{Bevis:2007qz}
\bibinfo{author}{\bibfnamefont{N.}~\bibnamefont{Bevis}},
  \bibinfo{author}{\bibfnamefont{M.}~\bibnamefont{Hindmarsh}},
  \bibinfo{author}{\bibfnamefont{M.}~\bibnamefont{Kunz}}, \bibnamefont{and}
  \bibinfo{author}{\bibfnamefont{J.}~\bibnamefont{Urrestilla}},
  \bibinfo{journal}{Phys. Rev.} \textbf{\bibinfo{volume}{D76}},
  \bibinfo{pages}{043005} (\bibinfo{year}{2007}{\natexlab{a}}),
  \eprint{arXiv:0704.3800 [astro-ph]}.

\bibitem[{\citenamefont{Pogosian and Wyman}(2007)}]{Pogosian:2007gi}
\bibinfo{author}{\bibfnamefont{L.}~\bibnamefont{Pogosian}} \bibnamefont{and}
  \bibinfo{author}{\bibfnamefont{M.}~\bibnamefont{Wyman}}
  (\bibinfo{year}{2007}), \eprint{arXiv:0711.0747 [astro-ph]}.

\bibitem[{\citenamefont{Copeland et~al.}(2004)\citenamefont{Copeland, Myers,
  and Polchinski}}]{Copeland:2003bj}
\bibinfo{author}{\bibfnamefont{E.~J.} \bibnamefont{Copeland}},
  \bibinfo{author}{\bibfnamefont{R.~C.} \bibnamefont{Myers}}, \bibnamefont{and}
  \bibinfo{author}{\bibfnamefont{J.}~\bibnamefont{Polchinski}},
  \bibinfo{journal}{JHEP} \textbf{\bibinfo{volume}{06}}, \bibinfo{pages}{013}
  (\bibinfo{year}{2004}), \eprint{hep-th/0312067}.

\bibitem[{\citenamefont{Polchinski}(2004)}]{Polchinski:2004ia}
\bibinfo{author}{\bibfnamefont{J.}~\bibnamefont{Polchinski}}
  (\bibinfo{year}{2004}), \eprint{hep-th/0412244}.

\bibitem[{\citenamefont{Sarangi and Tye}(2002)}]{Sarangi:2002yt}
\bibinfo{author}{\bibfnamefont{S.}~\bibnamefont{Sarangi}} \bibnamefont{and}
  \bibinfo{author}{\bibfnamefont{S.~H.~H.} \bibnamefont{Tye}},
  \bibinfo{journal}{Phys. Lett.} \textbf{\bibinfo{volume}{B536}},
  \bibinfo{pages}{185} (\bibinfo{year}{2002}), \eprint{hep-th/0204074}.

\bibitem[{\citenamefont{Jones et~al.}(2003)\citenamefont{Jones, Stoica, and
  Tye}}]{Jones:2003da}
\bibinfo{author}{\bibfnamefont{N.~T.} \bibnamefont{Jones}},
  \bibinfo{author}{\bibfnamefont{H.}~\bibnamefont{Stoica}}, \bibnamefont{and}
  \bibinfo{author}{\bibfnamefont{S.~H.~H.} \bibnamefont{Tye}},
  \bibinfo{journal}{Phys. Lett.} \textbf{\bibinfo{volume}{B563}},
  \bibinfo{pages}{6} (\bibinfo{year}{2003}), \eprint{hep-th/0303269}.

\bibitem[{\citenamefont{Vincent et~al.}(1997)\citenamefont{Vincent, Hindmarsh,
  and Sakellariadou}}]{Vincent:1996rb}
\bibinfo{author}{\bibfnamefont{G.~R.} \bibnamefont{Vincent}},
  \bibinfo{author}{\bibfnamefont{M.}~\bibnamefont{Hindmarsh}},
  \bibnamefont{and}
  \bibinfo{author}{\bibfnamefont{M.}~\bibnamefont{Sakellariadou}},
  \bibinfo{journal}{Phys. Rev.} \textbf{\bibinfo{volume}{D56}},
  \bibinfo{pages}{637} (\bibinfo{year}{1997}), \eprint{astro-ph/9612135}.

\bibitem[{\citenamefont{Vincent et~al.}(1998)\citenamefont{Vincent, Antunes,
  and Hindmarsh}}]{Vincent:1997cx}
\bibinfo{author}{\bibfnamefont{G.}~\bibnamefont{Vincent}},
  \bibinfo{author}{\bibfnamefont{N.~D.} \bibnamefont{Antunes}},
  \bibnamefont{and}
  \bibinfo{author}{\bibfnamefont{M.}~\bibnamefont{Hindmarsh}},
  \bibinfo{journal}{Phys. Rev. Lett.} \textbf{\bibinfo{volume}{80}},
  \bibinfo{pages}{2277} (\bibinfo{year}{1998}), \eprint{hep-ph/9708427}.

\bibitem[{\citenamefont{Siemens et~al.}(2002)\citenamefont{Siemens, Olum, and
  Vilenkin}}]{Siemens:2002dj}
\bibinfo{author}{\bibfnamefont{X.}~\bibnamefont{Siemens}},
  \bibinfo{author}{\bibfnamefont{K.~D.} \bibnamefont{Olum}}, \bibnamefont{and}
  \bibinfo{author}{\bibfnamefont{A.}~\bibnamefont{Vilenkin}},
  \bibinfo{journal}{Phys. Rev.} \textbf{\bibinfo{volume}{D66}},
  \bibinfo{pages}{043501} (\bibinfo{year}{2002}), \eprint{gr-qc/0203006}.

\bibitem[{\citenamefont{Martins and Shellard}(2006)}]{Martins:2005es}
\bibinfo{author}{\bibfnamefont{C.~J. A.~P.} \bibnamefont{Martins}}
  \bibnamefont{and} \bibinfo{author}{\bibfnamefont{E.~P.~S.}
  \bibnamefont{Shellard}}, \bibinfo{journal}{Phys. Rev.}
  \textbf{\bibinfo{volume}{D73}}, \bibinfo{pages}{043515}
  (\bibinfo{year}{2006}), \eprint{astro-ph/0511792}.

\bibitem[{\citenamefont{Ringeval et~al.}(2007)\citenamefont{Ringeval,
  Sakellariadou, and Bouchet}}]{Ringeval:2005kr}
\bibinfo{author}{\bibfnamefont{C.}~\bibnamefont{Ringeval}},
  \bibinfo{author}{\bibfnamefont{M.}~\bibnamefont{Sakellariadou}},
  \bibnamefont{and} \bibinfo{author}{\bibfnamefont{F.}~\bibnamefont{Bouchet}},
  \bibinfo{journal}{JCAP} \textbf{\bibinfo{volume}{0702}}, \bibinfo{pages}{023}
  (\bibinfo{year}{2007}), \eprint{astro-ph/0511646}.

\bibitem[{\citenamefont{Vanchurin et~al.}(2006)\citenamefont{Vanchurin, Olum,
  and Vilenkin}}]{Vanchurin:2005pa}
\bibinfo{author}{\bibfnamefont{V.}~\bibnamefont{Vanchurin}},
  \bibinfo{author}{\bibfnamefont{K.~D.} \bibnamefont{Olum}}, \bibnamefont{and}
  \bibinfo{author}{\bibfnamefont{A.}~\bibnamefont{Vilenkin}},
  \bibinfo{journal}{Phys. Rev.} \textbf{\bibinfo{volume}{D74}},
  \bibinfo{pages}{063527} (\bibinfo{year}{2006}), \eprint{gr-qc/0511159}.

\bibitem[{\citenamefont{Hindmarsh et~al.}()\citenamefont{Hindmarsh, Stuckey,
  and Bevis}}]{Hindmarsh:2008}
\bibinfo{author}{\bibfnamefont{M.}~\bibnamefont{Hindmarsh}},
  \bibinfo{author}{\bibfnamefont{S.}~\bibnamefont{Stuckey}}, \bibnamefont{and}
  \bibinfo{author}{\bibfnamefont{N.}~\bibnamefont{Bevis}}, \bibinfo{note}{in
  preparation}.

\bibitem[{\citenamefont{Moore et~al.}(2001)\citenamefont{Moore, Shellard, and
  Martins}}]{Moore:2001px}
\bibinfo{author}{\bibfnamefont{J.~N.} \bibnamefont{Moore}},
  \bibinfo{author}{\bibfnamefont{E.~P.~S.} \bibnamefont{Shellard}},
  \bibnamefont{and} \bibinfo{author}{\bibfnamefont{C.~J. A.~P.}
  \bibnamefont{Martins}}, \bibinfo{journal}{Phys. Rev.}
  \textbf{\bibinfo{volume}{D65}}, \bibinfo{pages}{023503}
  (\bibinfo{year}{2001}), \eprint{hep-ph/0107171}.

\bibitem[{\citenamefont{Copeland and Saffin}(2005)}]{Copeland:2005cy}
\bibinfo{author}{\bibfnamefont{E.~J.} \bibnamefont{Copeland}} \bibnamefont{and}
  \bibinfo{author}{\bibfnamefont{P.~M.} \bibnamefont{Saffin}},
  \bibinfo{journal}{JHEP} \textbf{\bibinfo{volume}{11}}, \bibinfo{pages}{023}
  (\bibinfo{year}{2005}), \eprint{hep-th/0505110}.

\bibitem[{\citenamefont{Hindmarsh and Saffin}(2006)}]{Hindmarsh:2006qn}
\bibinfo{author}{\bibfnamefont{M.}~\bibnamefont{Hindmarsh}} \bibnamefont{and}
  \bibinfo{author}{\bibfnamefont{P.~M.} \bibnamefont{Saffin}},
  \bibinfo{journal}{JHEP} \textbf{\bibinfo{volume}{08}}, \bibinfo{pages}{066}
  (\bibinfo{year}{2006}), \eprint{hep-th/0605014}.

\bibitem[{\citenamefont{Rajantie et~al.}(2007)\citenamefont{Rajantie,
  Sakellariadou, and Stoica}}]{Rajantie:2007hp}
\bibinfo{author}{\bibfnamefont{A.}~\bibnamefont{Rajantie}},
  \bibinfo{author}{\bibfnamefont{M.}~\bibnamefont{Sakellariadou}},
  \bibnamefont{and} \bibinfo{author}{\bibfnamefont{H.}~\bibnamefont{Stoica}},
  \bibinfo{journal}{JCAP} \textbf{\bibinfo{volume}{0711}}, \bibinfo{pages}{021}
  (\bibinfo{year}{2007}), \eprint{arXiv:0706.3662 [hep-th]}.

\bibitem[{\citenamefont{Urrestilla and Vilenkin}(2008)}]{Urrestilla:2007yw}
\bibinfo{author}{\bibfnamefont{J.}~\bibnamefont{Urrestilla}} \bibnamefont{and}
  \bibinfo{author}{\bibfnamefont{A.}~\bibnamefont{Vilenkin}},
  \bibinfo{journal}{JHEP} \textbf{\bibinfo{volume}{02}}, \bibinfo{pages}{037}
  (\bibinfo{year}{2008}), \eprint{arXiv:0712.1146 [hep-th]}.

\bibitem[{\citenamefont{Tye et~al.}(2005)\citenamefont{Tye, Wasserman, and
  Wyman}}]{Tye:2005fn}
\bibinfo{author}{\bibfnamefont{S.~H.~H.} \bibnamefont{Tye}},
  \bibinfo{author}{\bibfnamefont{I.}~\bibnamefont{Wasserman}},
  \bibnamefont{and} \bibinfo{author}{\bibfnamefont{M.}~\bibnamefont{Wyman}},
  \bibinfo{journal}{Phys. Rev.} \textbf{\bibinfo{volume}{D71}},
  \bibinfo{pages}{103508} (\bibinfo{year}{2005}), \eprint{astro-ph/0503506}.

\bibitem[{\citenamefont{Avgoustidis and Shellard}(2007)}]{Avgoustidis:2007aa}
\bibinfo{author}{\bibfnamefont{A.}~\bibnamefont{Avgoustidis}} \bibnamefont{and}
  \bibinfo{author}{\bibfnamefont{E.~P.~S.} \bibnamefont{Shellard}}
  (\bibinfo{year}{2007}), \eprint{arXiv:0705.3395 [astro-ph]}.

\bibitem[{\citenamefont{Copeland et~al.}(2006)\citenamefont{Copeland, Kibble,
  and Steer}}]{Copeland:2006eh}
\bibinfo{author}{\bibfnamefont{E.~J.} \bibnamefont{Copeland}},
  \bibinfo{author}{\bibfnamefont{T.~W.~B.} \bibnamefont{Kibble}},
  \bibnamefont{and} \bibinfo{author}{\bibfnamefont{D.~A.} \bibnamefont{Steer}},
  \bibinfo{journal}{Phys. Rev. Lett.} \textbf{\bibinfo{volume}{97}},
  \bibinfo{pages}{021602} (\bibinfo{year}{2006}), \eprint{hep-th/0601153}.

\bibitem[{\citenamefont{Copeland
  et~al.}(2007{\natexlab{a}})\citenamefont{Copeland, Kibble, and
  Steer}}]{Copeland:2006if}
\bibinfo{author}{\bibfnamefont{E.~J.} \bibnamefont{Copeland}},
  \bibinfo{author}{\bibfnamefont{T.~W.~B.} \bibnamefont{Kibble}},
  \bibnamefont{and} \bibinfo{author}{\bibfnamefont{D.~A.} \bibnamefont{Steer}},
  \bibinfo{journal}{Phys. Rev.} \textbf{\bibinfo{volume}{D75}},
  \bibinfo{pages}{065024} (\bibinfo{year}{2007}{\natexlab{a}}),
  \eprint{hep-th/0611243}.

\bibitem[{\citenamefont{Saffin}(2005)}]{Saffin:2005cs}
\bibinfo{author}{\bibfnamefont{P.~M.} \bibnamefont{Saffin}},
  \bibinfo{journal}{JHEP} \textbf{\bibinfo{volume}{09}}, \bibinfo{pages}{011}
  (\bibinfo{year}{2005}), \eprint{hep-th/0506138}.

\bibitem[{\citenamefont{Bettencourt et~al.}(1997)\citenamefont{Bettencourt,
  Laguna, and Matzner}}]{Bettencourt:1996qe}
\bibinfo{author}{\bibfnamefont{L.~M.~A.} \bibnamefont{Bettencourt}},
  \bibinfo{author}{\bibfnamefont{P.}~\bibnamefont{Laguna}}, \bibnamefont{and}
  \bibinfo{author}{\bibfnamefont{R.~A.} \bibnamefont{Matzner}},
  \bibinfo{journal}{Phys. Rev. Lett.} \textbf{\bibinfo{volume}{78}},
  \bibinfo{pages}{2066} (\bibinfo{year}{1997}), \eprint{hep-ph/9612350}.

\bibitem[{\citenamefont{Salmi et~al.}(2007)}]{Salmi:2007ah}
\bibinfo{author}{\bibfnamefont{P.}~\bibnamefont{Salmi}} \bibnamefont{et~al.}
  (\bibinfo{year}{2007}), \eprint{arXiv:0712.1204 [hep-th]}.

\bibitem[{\citenamefont{Cui et~al.}(2008)\citenamefont{Cui, Martin, Morrissey,
  and Wells}}]{Cui:2007js}
\bibinfo{author}{\bibfnamefont{Y.}~\bibnamefont{Cui}},
  \bibinfo{author}{\bibfnamefont{S.~P.} \bibnamefont{Martin}},
  \bibinfo{author}{\bibfnamefont{D.~E.} \bibnamefont{Morrissey}},
  \bibnamefont{and} \bibinfo{author}{\bibfnamefont{J.~D.} \bibnamefont{Wells}},
  \bibinfo{journal}{Phys. Rev.} \textbf{\bibinfo{volume}{D77}},
  \bibinfo{pages}{043528} (\bibinfo{year}{2008}), \eprint{0709.0950}.

\bibitem[{\citenamefont{Copeland
  et~al.}(2007{\natexlab{b}})\citenamefont{Copeland, Firouzjahi, Kibble, and
  Steer}}]{Copeland:2007nv}
\bibinfo{author}{\bibfnamefont{E.~J.} \bibnamefont{Copeland}},
  \bibinfo{author}{\bibfnamefont{H.}~\bibnamefont{Firouzjahi}},
  \bibinfo{author}{\bibfnamefont{T.~W.~B.} \bibnamefont{Kibble}},
  \bibnamefont{and} \bibinfo{author}{\bibfnamefont{D.~A.} \bibnamefont{Steer}}
  (\bibinfo{year}{2007}{\natexlab{b}}), \eprint{arXiv:0712.0808 [hep-th]}.

\bibitem[{\citenamefont{Nielsen and Olesen}(1973)}]{Nielsen:1973cs}
\bibinfo{author}{\bibfnamefont{H.~B.} \bibnamefont{Nielsen}} \bibnamefont{and}
  \bibinfo{author}{\bibfnamefont{P.}~\bibnamefont{Olesen}},
  \bibinfo{journal}{Nucl. Phys.} \textbf{\bibinfo{volume}{B61}},
  \bibinfo{pages}{45} (\bibinfo{year}{1973}).

\bibitem[{\citenamefont{Bogomol'nyi}(1976)}]{Bogomolnyi:1976}
\bibinfo{author}{\bibfnamefont{E.}~\bibnamefont{Bogomol'nyi}},
  \bibinfo{journal}{Sov. J. Nucl. Phys.} \textbf{\bibinfo{volume}{24}},
  \bibinfo{pages}{449} (\bibinfo{year}{1976}).

\bibitem[{\citenamefont{Moriarty et~al.}(1988)\citenamefont{Moriarty, Myers,
  and Rebbi}}]{Moriarty:1988fx}
\bibinfo{author}{\bibfnamefont{K.~J.~M.} \bibnamefont{Moriarty}},
  \bibinfo{author}{\bibfnamefont{E.}~\bibnamefont{Myers}}, \bibnamefont{and}
  \bibinfo{author}{\bibfnamefont{C.}~\bibnamefont{Rebbi}},
  \bibinfo{journal}{Phys. Lett.} \textbf{\bibinfo{volume}{B207}},
  \bibinfo{pages}{411} (\bibinfo{year}{1988}).

\bibitem[{\citenamefont{Bevis et~al.}(2007{\natexlab{b}})\citenamefont{Bevis,
  Hindmarsh, Kunz, and Urrestilla}}]{Bevis:2006mj}
\bibinfo{author}{\bibfnamefont{N.}~\bibnamefont{Bevis}},
  \bibinfo{author}{\bibfnamefont{M.}~\bibnamefont{Hindmarsh}},
  \bibinfo{author}{\bibfnamefont{M.}~\bibnamefont{Kunz}}, \bibnamefont{and}
  \bibinfo{author}{\bibfnamefont{J.}~\bibnamefont{Urrestilla}},
  \bibinfo{journal}{Phys. Rev.} \textbf{\bibinfo{volume}{D75}},
  \bibinfo{pages}{065015} (\bibinfo{year}{2007}{\natexlab{b}}),
  \eprint{astro-ph/0605018}.

\bibitem[{\citenamefont{Bevis and Hindmarsh}()}]{LATfield}
\bibinfo{author}{\bibfnamefont{N.}~\bibnamefont{Bevis}} \bibnamefont{and}
  \bibinfo{author}{\bibfnamefont{M.}~\bibnamefont{Hindmarsh}},
  \urlprefix\url{www.latfield.org}.

\bibitem[{Cos()}]{Cosmos}
\bibinfo{note}{UK National Cosmology Supercomputer},
  \urlprefix\url{www.damtp.cam.ac.uk/cosmos/}.

\bibitem[{\citenamefont{{Matzner}}(1988)}]{Matzner:1988}
\bibinfo{author}{\bibfnamefont{R.~A.} \bibnamefont{{Matzner}}},
  \bibinfo{journal}{Computers in Physics} \textbf{\bibinfo{volume}{2}},
  \bibinfo{pages}{51} (\bibinfo{year}{1988}).

\bibitem[{\citenamefont{Laguna and Matzner}(1990)}]{Laguna:1990it}
\bibinfo{author}{\bibfnamefont{P.}~\bibnamefont{Laguna}} \bibnamefont{and}
  \bibinfo{author}{\bibfnamefont{R.~A.} \bibnamefont{Matzner}},
  \bibinfo{journal}{Phys. Rev.} \textbf{\bibinfo{volume}{D41}},
  \bibinfo{pages}{1751} (\bibinfo{year}{1990}).

\bibitem[{\citenamefont{Shellard}(1987)}]{Shellard:1987bv}
\bibinfo{author}{\bibfnamefont{E.~P.~S.} \bibnamefont{Shellard}},
  \bibinfo{journal}{Nucl. Phys.} \textbf{\bibinfo{volume}{B283}},
  \bibinfo{pages}{624} (\bibinfo{year}{1987}).

\bibitem[{\citenamefont{Kajantie et~al.}(1998)\citenamefont{Kajantie,
  Karjalainen, Laine, Peisa, and Rajantie}}]{Kajantie:1998bg}
\bibinfo{author}{\bibfnamefont{K.}~\bibnamefont{Kajantie}},
  \bibinfo{author}{\bibfnamefont{M.}~\bibnamefont{Karjalainen}},
  \bibinfo{author}{\bibfnamefont{M.}~\bibnamefont{Laine}},
  \bibinfo{author}{\bibfnamefont{J.}~\bibnamefont{Peisa}}, \bibnamefont{and}
  \bibinfo{author}{\bibfnamefont{A.}~\bibnamefont{Rajantie}},
  \bibinfo{journal}{Phys. Lett.} \textbf{\bibinfo{volume}{B428}},
  \bibinfo{pages}{334} (\bibinfo{year}{1998}), \eprint{hep-ph/9803367}.

\end{thebibliography}

\end{document}